\documentclass[fleqn,usenatbib]{mnras}

\usepackage{newtxtext,newtxmath}
\usepackage[modulo]{lineno}
\usepackage{ulem}
\usepackage{soul,xcolor}
\usepackage{xcolor}

\newcommand{\teff}   {$T_{\rm eff}$} 
\newcommand{\tirfm}  {$T_{\rm IRFM}$}
 
\newcommand{\logg}   {$\log g$} 

\newcommand{\ali}    {A(Li)}

\newcommand{\V}      {$V$}
\newcommand{\I}      {$I$}
\newcommand{\J}      {$J$}
\newcommand{\K}      {$K$}
\newcommand{\g}      {$g$}
\newcommand{\ii}     {$i$}
\newcommand{\vi}     {$V$--$I$}
\newcommand{\vj}     {$V$--$J$}
\newcommand{\vk}     {$V$--$K$}
\newcommand{\ebv}    {E($B$--$V$)}
\newcommand{\gizero} {$(g-i)$$_{\rm 0}$} 
\newcommand{\gi}     {$(g-i)$} 
\newcommand{\simlt}  {\lower.5ex\hbox{\ltsima}} 
\newcommand{\simgt}  {\lower.5ex\hbox{\gtsima}}
\newcommand{\msun}   {\rm M$_\odot$} 
\newcommand{\ltsima} {$\; \buildrel < \over \sim \;$}
\newcommand{\gtsima} {$\; \buildrel > \over \sim \;$}

\usepackage[T1]{fontenc}

\DeclareRobustCommand{\VAN}[3]{#2}
\let\VANthebibliography\thebibliography
\def\thebibliography{\DeclareRobustCommand{\VAN}[3]{##3}\VANthebibliography}

\usepackage{graphicx}	
\usepackage{amsmath}	

\usepackage{longtable}

\title[The Astration of Primordial Lithium]{A Critique of the Spite Plateau, and the Astration of Primordial Lithium}

\author[J. E. Norris et al.]{
J. E. Norris,$^{1}$\thanks{E-mail: john.norris@anu.edu.au}
D. Yong,$^{1,2}$
A. Frebel,$^{3}$
S.G. Ryan$^{4}$
\\
$^{1}$Research School of Astronomy and Astrophysics, Australian National University, Canberra, ACT 0200, Australia\\
$^{2}$ARC Centre of Excellence for Astrophysics in Three Dimensions (ASTRO-3D), Australia\\
$^{3}$Department of Physics and Kavli Institute for Astrophysics and Space Research, Massachusetts Institute of Technology, Cambridge, MA 02139, USA\\
$^{4}$Department of Physics, Astronomy and Mathematics, University of Hertfordshire, College Lane, Hatfield AL10 9AB, UK\\
}

\date{Accepted 2023 March 10. Received 2023 March 06; in original form 2022 December 03} 
\pubyear{2022}

\begin{document}
\setstcolor{red}
\label{firstpage}
\pagerange{\pageref{firstpage}--\pageref{lastpage}}
\maketitle

\begin{abstract}
We investigate the distribution of the lithium abundances, {\ali}, of metal-poor dwarf and subgiant stars within the limits 5500~K $<$ {\teff} $<$ 6700~K, --6.0 $<$ [Fe/H] $<$ --1.5, and {\logg} {\simgt} 3.5 (a superset of parameters first adopted by Spite and Spite), using literature data for some 200 stars.  We address the problem of the several methods that yield {\teff} differences up to 350~K, and hence uncertainties of 0.3 dex in [Fe/H] and {\ali}, by anchoring {\teff} to the Infrared Flux Method.  We seek to understand the behaviour of {\ali} as a function of [Fe/H] -- small dispersion at highest [Fe/H], ``meltdown'' at intermediate values (i.e. large spread in Li below the Spite Plateau), and extreme variations at lowest [Fe/H].  Decreasing {\ali} is accompanied by increasing dispersion. Insofar as [Fe/H] increases as the universe ages, the behavior of {\ali} reflects chaotic star formation involving destruction of primordial Li, which settles to the classic Spite Plateau, with {\ali} $\sim$ 2.3, by the time the Galactic halo reaches [Fe/H] $\sim$ --3.0.  We consider three phases: (1) first star formation in C-rich environments ([C/Fe] $>$ 2.3), with depleted Li; (2) silicates-dominated star formation and destruction of primordial Li during pre-main-sequence evolution; and (3) materials from these two phases co-existing and coalescing to form C-rich stars with {\ali} below the Spite Plateau, leading to a toy model with the potential to explain the ``meltdown''.  We comment on the results of Mucciarelli et al. on the Lower RGB, and the suggestion of Aguado et al. favouring a lower primordial lithium abundance than generally accepted.

\end{abstract}



\color{black}
\section{Introduction} \label{sec:intro}

One of the basic predictions of the Big Bang cosmological model is that lithium is produced in a hot Big Bang \citep{wagoner67, wagoner73}. Subsequently, \citet{spite82} first reported the small dispersion in lithium abundances in near-main-sequence stars in the ranges 5500~K $<$ {\teff} $<$ 6500~K and --2.4 $<$ [Fe/H]\footnote{Throughout this paper, stellar chemical abundances assume 1D,LTE (one-dimension, local-thermodynamic-equilibrium) modelling, unless otherwise stated.}  $<$ --1.1 (later to be known as the Spite Plateau), and proposed these stars as potential probes of the Big Bang prediction.  They reported a lithium abundance of the Spite Plateau of {\ali} = 2.05 $\pm$ 0.15; but also noted: ``The observed lithium abundance is thus at least a lower limit of the $^{7}$Li produced by the Big Bang''.  Two decades later, after considerable further investigation of the properties of the Big Bang and the Spite Plateau, by many investigators, it became clear there is indeed a significant difference between theory and observation as documented, for example, by \citet{cyburt08}, who reported a predicted primordial lithium abundance (based on WMAP \citep{dunkley09} and Big Bang $\Lambda$CDM cosmology)\footnote{We note the \citet{planck20} result: ``We do not discuss other light elements, such as ... lithium, since the observed abundance measurements and their interpretation in terms of the standard models of BBN are more controversial (see \citet{fields11, fields14}, for reviews). The Planck results do not shed any further light on these problems compared to earlier CMB experiments.''} of {\ali}$_{\rm P}$ = 2.72 $\pm$ 0.06 and the Spite Plateau value of {\ali} = 2.09 $\pm$ 0.16 for near-main-sequence stars $-$ a difference of 0.63 $\pm$ 0.17 dex.  This was referred to as ``The Lithium Problem''.

The situation, however, became considerably more challenging as it became clear that many dwarfs of lower metallicity, in the range --6.0 $<$[Fe/H] $<$ --2.5 (and ages most likely closer to that of the Big Bang), have lithium abundances considerably smaller than {\ali} = 2.0.  The majority of the seven known dwarfs with [Fe/H] $<$ --4.5 have 2.0 $<$ {\ali} $<$ 0.5 (beginning with HE~1327--2326 which has [Fe/H] = --5.6 and {\ali} $<$ 1.6 \citep{frebel05}; and in the range --4.0 $<$ [Fe/H] $<$ --3.0, \citet{sbordone10} and \citet{bonifacio12} reported a ``meltdown'' of the Spite Plateau, once again involving near-main-sequence stars with {\ali} values considerably less than 2.0.

Table~\ref{tab:tab1} presents a brief list of some 20 important milestones (including those above) relevant to the investigations of Li abundances, which have played an important role in our current understanding of the early universe.

 \begin{table*}
	\caption{Major Milestones in the Study of Lithium Abundances in the Metal-Poor Near-Main-Sequence Stars}
	\label{tab:tab1}
	\begin{tabular}{ll}
	\hline
        Milestone   &  
        Authors$^1$  \\
        \hline
Prediction of lithium production in the early stages of a homogeneous and isotropic expanding Universe  & WA67, WA73 \\
 & \\
Discovery of the Spite Plateau: $\langle${\ali}$\rangle$ = 2.05, for stars with 5500~K $<$ {\teff} $<$ 6250~K, &  SP82 \\
{\logg} $\simgt$ 3.5, and  --2.4 $<$ [Fe/H] $<$ --1.1  & \\ 
 & \\
9\% of stars lie below the Plateau in the range 1.1 $<$ {\ali} $<$ 2.0 ({\teff} $>$ 5800~K, --3.8 $<$ [Fe/H] $<$ --1.7) &  TH94 \\
 & \\
The existence of ``rare cases of well-observed stars (with similar {\teff} and [Fe/H] $\sim$ --3.0) which cannot have  &  RY96 \\
the same Li abundance'' & \\
 & \\
Spectra with S/N $>$ 100 reveal the Spite Plateau (on the limited range 6100~K $<$ {\teff} $<$ 6300~K) is slightly & RY99 \\
  inclined with respect to [Fe/H] & \\
 & \\
Suggestion that ``ultra-Li-depleted halo stars and blue stragglers are manifestations of the same phenomenon'' & RY01a, RY02 \\ 
 & \\
Discovery of hyper-metal-poor (HMP) subgiant HE~1327--2326 with [Fe/H]$_{\rm 1D,LTE}$ = --5.66 and {\ali}$_{\rm 1D,LTE}$ $<$ 0.70 & FR05, AO06, FR08 \\
 & \\
Stellar model atmosphere computations establish {\ali}$_{\rm 3D,NLTE}$ and {\ali}$_{\rm 1D,LTE}$ values are essentially the same & AS03 \\
& \\
Spectra with S/N $>$ 300 yield Spite Plateau (5800~K $<$ {\teff} $<$ 6400~K) slightly inclined over --2.9 $<$ [Fe/H] $<$ --1.0  & AS06 \\
& \\
The Spite Plateau $\langle${\ali}$\rangle$ = 2.09 is lower than the WMAP and Big Bang $\Lambda$CDM primordial value {\ali}$_{\rm P}$ = 2.72    & CY08 \\
& \\
Theoretical analysis of sensitivity of stellar atmospheric lithium abundances to standard stellar evolution modelling & DE90 \\
& \\
Important theoretical suggestions to explain lower observed Spite Plateau {\ali} value: (a) Gravitational settling & RI05,  PI06, FU15 \\
 in the presence of weak turbulence in low-mass main sequence stars; (b) Depletion of Li by a first generation of stars; &  \\
(c) Convective overshoot and residual mass accretion during pre-main-sequence and main-sequence evolution &  \\
& \\
The cooler {\teff} limit of the Plateau moves to a higher value as [Fe/H] decreases & ME10 \\
& \\
The components of the double-line binary CS22876-032 have different {\ali} = 2.22 and 1.75 ({\teff} = 6500, 5900~K & TH94, GO08 \\ 
and [Fe/H] = --3.66, --3.57) & \\
& \\
Meltdown: large {\ali} spreads below the Spite Plateau value observed for [Fe/H] {\simlt} --3.0 & SB10, BO12, 15, 18 \\
& \\
Discovery of most Fe-poor star SM~0313--6708$^2$, with [Fe/H] $<$ --7.3 $-$ a red giant deduced to have had & KE14, FR15 \\
main sequence {\ali}$_{\rm MS}$ $\sim$~2.0 & \\
& \\
Spectra with S/N $>$ 40--100 for 7 stars with {\teff} $>$ 5500~K, {\logg} $>$ 3.5, and [Fe/H] $<$ --3.5 yield  & MA17a \\ 
$\langle${\ali}$\rangle$ = 1.90, with $\sigma$=0.10  &   \\
& \\
At least 12 of the 14 stars known in 2019 to have [Fe/H] $<$ --4.5 are C-rich, with [C/Fe] $\ge$ 0.7.  11 of them & NO19 \\
have [C/Fe] $>$ 3.0  &  \\
& \\
For the 8 stars having [Fe/H] $<$ --4.5, {\logg} $\simgt$ 3.5, and lithium estimates, the mean lithium abundance is & FR08,19, AG19,\\
$\langle${\ali}$\rangle$ $\leq$ 1.5 & BO15,18, CA11,16, \\
& HA15, ST18 \\ 
Suggestion that C enhancement and Li depletion are not directly related; 2 C-rich stars at [Fe/H] $\sim$ --3.0 & MA17b \\
have normal {\ali} & \\ 
& \\
Lithium upper envelope over the range --6.0 $<$ [Fe/H] $<$ --2.5 ``nearly constant'' at {\ali} $\sim$~2.0 suggests & AG19 \\
 ``lower primordial production''  & \\
& \\
Discovery of a thin lithium plateau among metal-poor red giant branch stars & MU22 \\ 
& \\
\hline
	\end{tabular}
        \\
        $^1${WA67} = \citet{wagoner67}, {WA73} = \citet{wagoner67}, SP82  = \citet{spite82}, TH94  = \citet{thorburn94}, 
        RY96  = \citet{ryan96}, RY99  = \citet{ryan99}, RY01a = \citet{ryan01a}, RY02  = \citet{ryan02}, FR05  = \citet{frebel05}, AS06  = \citet{asplund06}, FR08  = \citet{frebel08}, AS03  = \citet{asplund03}, CY08  = \citet{cyburt08} (WMAP = Wilkinson Microwave Anisotropy Probe), DE90  = \citet{del90}, RI05  = \citet{ric05}, PI06  = \citet{pia06}, FU15  = \citet{fu15}, ME10  = \citet{melendez10}, GO08  = \citet{gonzalez08}, SB10  = \citet{sbordone10}, BO12  = \citet{bonifacio12}, BO15  = \citet{bonifacio15}, BO18  = \citet{bonifacio18}, KE14  = \citet{keller14}, FR15 = \citet{frebel15b}, MA17a = \citet{matsuno17a}, NO19  = \citet[see Table~6]{norris19}, FR19 = \citet{frebel19}, AG19  = \citet{aguado19}, CA11  = \citet{caffau11}, CA16  = \citet{caffau16}, HA15  = \citet{hansen15}, ST18  = \citet{starkenburg18}, MA17b = \citet{matsuno17b}, MU22 = \citet{mucciarelli22} \\
$^2$ {SM~0313--6708 = SMSS~J031300.36--670839.3}

\end{table*}
\newpage   

\begin{figure*}
    \includegraphics[width=.5\hsize]{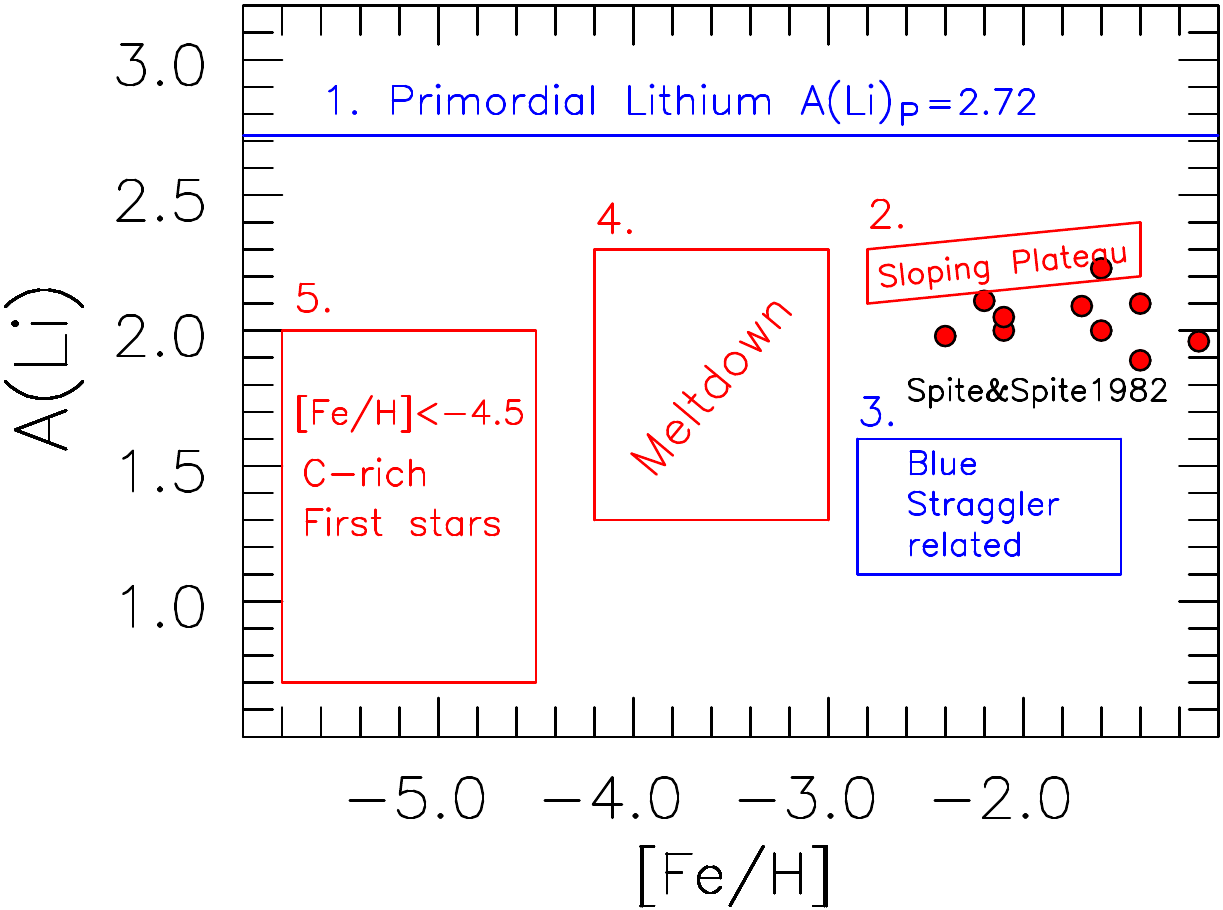}
    \caption{The five lithium problems. Red filled circles represent
      the original discovery data of \citet{spite82}.}
    \label{fig:liprob}
\end{figure*}

The aim of the present work is threefold.  First, we present a critique of the lithium abundances of near-main-sequence turnoff stars on the Spite Plateau \citep{spite82} in the range --2.4 $<$ [Fe/H] $<$ --1.1, together with the observed followup which stretches down to [Fe/H] = --6.0.  By way of closer introduction, Figure 1 provides the background for discussion of what we see as five (not one) lithium problems.  These are: \newline (1) the primordial ``Lithium Problem'' discussed above $-$ {\ali}$_{\rm P}$ = 2.72, while {\ali}$_{\rm Spite Plateau}$ = 2.09 \citep{cyburt08}; \newline(2) a sloping plateau \citep{ryan99, asplund06}; \newline(3) a group of stars with very low {\ali} values which exhibit relatively rapid rotation and appear to be related to ``blue stragglers'' \citep{thorburn94, ryan01b, ryan02}; \newline(4) the ``meltdown'' of the plateau \citep{sbordone10, bonifacio12, bonifacio15, bonifacio18} referred to above; and \newline(5) the enormous range in {\ali} ($<$0.5 $<$ {\ali} $<$ 2.0) among the most Fe-poor stars, with (in at least some cases) an anticorrelation between A(Li) and carbon abundance, [C/Fe]\footnote{We note that this paper does not consider the isotopic ratios of Li (on which there is also considerable literature debate).}.

Second, having collated a literature database of {\ali} values, we correct them to a common temperature scale, by taking into account differences resulting from the various techniques adopted by authors -- with a view to improving their relative accuracy.  We also determine a further data set based on {\it ab initio} Infrared Flux Method (IRFM) temperatures, determined for stars having appropriate literature colours.

Third, we examine the various physical processes noted in Table~\ref{tab:tab1} that are potentially responsible for the ``problems'' summarized in Figure~\ref{fig:liprob} for {\ali} values below that of the primordial value $-$ that is, the phenomena responsible for the ``astration'' of the primordial lithium.  In particular, we follow the suggestions of \citet{bromm03}, \citet{frebel07}, and \citet{chiaki17}, and references therein, concerning star formation in a C-rich environment to explain the stars with [Fe/H] $<$ --4.5; and of \citet{fu15} on the astration of Li in C-normal stars in the range --3.2 $<$ [M/H] $<$ --1.5 during their pre-main-sequence phase.  In the range --4.5 $<$ [Fe/H] $<$ --3.0, we then follow the suggestions of \citet{norris13} and \citet{norris19} to investigate further the toy model they used to explain the fact that 20--30\% of stars with [Fe/H] $<$ --2.0 are CEMP-no stars\footnote{Carbon-Enhanced-Metal-Poor (C-rich) stars with [C/Fe] $>$ 0.7 and [Ba/Fe] $<$ 0.0 \color{blue}(\citealt{beers05}, \citealt{aoki07})}.  We shall argue that this model has the potential to explain the \citet{sbordone10}, \citet{bonifacio12, bonifacio15, bonifacio18} ``meltdown'' of the Spite Plateau\footnote{We draw the reader's attention to the fact that in Table 1 we identified several other potential causes of observed lithium abundances below the Spite Plateau $-$ in particular basic stellar evolutionary effects, {\it e.g.}  diffusion \citep{del90}; turbulent mixing within the stars themselves \citep{ric05}; and universal destruction of lithium by a first generation of stars following the Big Bang \citep{pia06}.  Our only justification for this is that they are not necessary in the suggestions we shall make.}.  In the Appendix we present a comparison between our conclusions with those presented by \citet{mucciarelli22} for stars on the Lower Red Giant Branch.

\section{Observational Data}

All lithium abundances in the present work are based on the Li I 6707~{\AA} doublet.  For our purposes we compiled from the literature a catalogue of near-main-sequence stars having {\teff}, {\logg}, and [Fe/H] data in the ranges 5500~K $<$ {\teff} $<$ 6720~K, {\logg} {\simgt}~3.5, and [Fe/H] $<$ --1.5.  As well as {\teff}, {\logg}, and [Fe/H] we catalogued A(Li) and/or equivalent widths of the Li~I doublet.  We excluded CEMP-s stars given that their carbon and lithium abundances have been modified as the result of binary stellar evolution effects, together with CEMP-r/s stars\footnote{See \citet{beers05}, \citet{lucatello05}, and \citet{aoki07} for details concerning these stars.}.

As will be discussed in more detail below, {\ali} values depend significantly on the different methods adopted by authors to determine {\teff} values.  We chose data samples to be as large as possible with a view to determining systemic differences between the effective temperatures of the various sources.  This initially resulted in 11 data subsets, each from an individual paper or from multiple papers by closely related co-workers.  In all but one case we accepted results based on observations original to the authors.  In an approximate time sequence, the 11 subsets (and their numbers of stars) are: \citet{thorburn94} (76 stars), \citet{ryan99, ryan01a, ryan01b} (31), \citet{fulbright00} (20), \citet{ford02} (10), \citet{asplund06} (18), Aoki and co-workers \citep{aoki09, aoki12, matsuno17a, matsuno17b, li15} (23), \citet{sbordone10}\footnote{Throughout the present work we adopt the {\ali}$_{\rm 1D,LTE}$, IRFM-based results in Table 4 of \citet{sbordone10}, in preference to the results therein based on other methods.} (29), \citet[and references therein]{melendez10}\footnote{We utilize the {\ali}$_{\rm 1D,LTE}$ values in Table 1 of \citet{melendez10}.}  (62), \citet{bonifacio12, bonifacio15, bonifacio18} (25), \citet{roederer14} (127), Reggiani and co-workers \citep{placco16, reggiani17} (24).  These sources provide the lithium and iron abundances necessary for our investigation.  To the best of our knowledge, all of the values we have used are based on 1D,LTE analyses.

\begin{figure*}
    \includegraphics[width=.85\hsize]{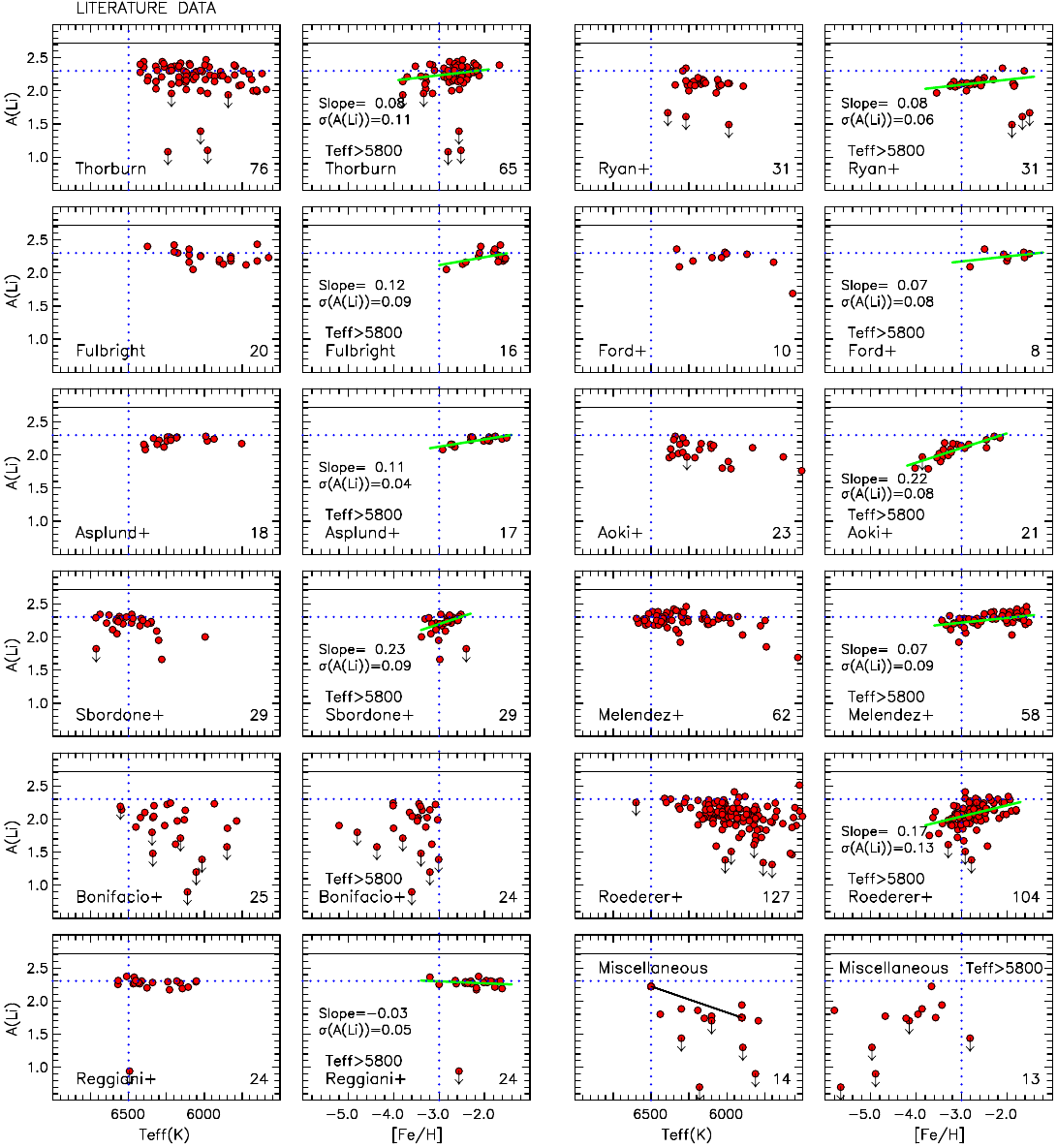}
    \caption{\small {{\ali} vs. {\teff} and [Fe/H] for the 12
        ``Literature'' data samples.  The black horizontal lines
        represent Primordial Lithium ({\ali}$_{\rm P}$ = 2.72).  The
        legends within each panel identify the sample and (at
        bottom-right) the number of points in the panel. In panels
        with [Fe/H] abscissae, the sloping line of best fit is shown
        (where appropriate) for stars in the range --3.5 {\simlt}
        [Fe/H] {\simlt} --1.5, for which the slope and RMS scatter are
        also presented; the adopted {\teff} lower limit is included.
        The co-joined points represent the EMP double-lined
        spectroscopic binary CS~22876-032. The dotted blue lines are
        the same in all panels and included to facilitate comparison
        between the data sets.  See text for discussion.}}
    \label{fig:authors1}
\end{figure*}

The above exceptional data subset is that of \citet{melendez10}, which adopts material from several sources in the literature to produce high-quality IRFM {\teff} values, which we shall use below to ``correct'' the literature {\teff} values of other sources to the IRFM scale.

These 11 subsets were augmented by a twelfth which comprises 14 ``miscellaneous'' near-main-sequence stars, with --6.0 $<$ [Fe/H] $<$ --2.8. Five of these have detected carbon abundances in the range 2.6 $<$ [C/Fe] $<$ 4.2, while a further six have limits between [C/Fe] $<$ 1.3 and [C/Fe] $<$ 2.0.  The sources of these data are \citet{hansen15}, \citet{frebel08, frebel19}, \citet{spite13}, \citet{aguado19}, \citet{caffau12, caffau16}, \citet{starkenburg18}, and \citet{lardo21}.

\section{Correcting the Available {\teff}, [Fe/H], and {\ali} Values} \label{sec:critique}

We examined the data for these stars in three ways.  First, we adopted the basic data from the literature; second, we transformed the {\teff}, [Fe/H], and {\ali} literature values onto the IRFM {\teff} scale by using the {\tirfm} based values of \citet{melendez10}; and third, we determined IRFM {\teff} values {\it{ab initio}} based on broadband infrared colours, complemented by {\gi} observations, as available.

\subsection{Step 1 - The Literature Data}\label{sec:literature}

Figure~\ref{fig:authors1} presents the dependence of {\ali} as a function of {\teff} and [Fe/H] for the 12 data sets for stars initially chosen to have {\teff} $>$ 5500~K and [Fe/H] $<$ --1.50.  In what follows we shall refer to this as the ``Literature'' sample. Also shown in the figure is the slope of the linear line of best fit in the ({\ali}, [Fe/H]) -- plane in the range --3.5 {\simlt} [Fe/H] {\simlt} --1.5 $-$ which has been discussed at some length by several workers {\color{blue}({\it e.g.} \citealt{thorburn94}, \citealt{ryan99})}, together with the RMS dispersion of the data points about the lines of best fit (excluding stars with only limiting values)\footnote{We note for completeness that we have excluded the 3$\sigma$ outlier (CS~22188-033) when determining the line of best fit for the \citet{sbordone10} data.}.  Setting aside the subset of ``miscellaneous'' stars having [Fe/H] $<$ --2.8 (in the bottom right panel of the figure) we comment on two basic aspects of the other 11 data sets.

\begin{table*}
	\caption{Mean Differences Relative to \citet{melendez10}, and {\ali} vs. [Fe/H] Slopes}
	\label{tab:tab2}
	\begin{tabular}{llrrrrccr}
	\hline
        Authors                             & 
        {\teff} Source                      & 
        $\langle\Delta${\teff}$\rangle$$^1$ &                
        $\langle\Delta$[Fe/H]$\rangle$$^2$  &               
        $\langle\Delta${\ali}$\rangle$$^3$  &              
        N$^4$                               &             
        Slope$^5$                           &            
        RMS$^6$                             &           
        N$^4$                               \\      
        (1) & (2) & (3) & (4) & (5) & (6) & (7) & (8) & (9) \\
        \hline
Thorburn (1994)               & Colours                   &  --192  &  --0.21  &   0.03 &  24 & 0.07  & 0.11  &  58   \\ 
Ryan et al.$^7$               & Colours \& H lines        &  --249  &  --0.17  & --0.15 &  16 & 0.06  & 0.06  &  27   \\ 
Fulbright (2000)              & EPM$^8$                   &  --211  &  --0.06  & --0.04 &  15 & 0.16  & 0.09  &  14   \\
Ford et al. (2002)            & IRFM                      &  --142  &  --0.12  & --0.06 &   6 & 0.10  & 0.10  &   6   \\ 
Asplund et al. (2006)         & H$\alpha$ line            &  --124  &    0.03  & --0.08 &  14 & 0.11  & 0.04  &  17   \\
Aoki et al.$^9$               & H lines \& colours        &  --219  &  --0.07  & --0.12 &   4 & 0.22  & 0.08  &  19   \\ 
Sbordone et al. (2010)        & IRFM                      &     38  &    0.07  &   0.04 &  13 & 0.18  & 0.09  &  26   \\ 
Melendez et al. (2010)        & IRFM                      &    ...  &     ...  &    ... & ... & 0.08  & 0.08  &  54   \\ 
Bonifacio et al.$^{10}$       & Colours                    &  --204  &  --0.20  & --0.16 &  25 &  ...  &  ...  &  ..   \\
Roederer et al. (2014)        & EPM$^8$                   &  --374  &  --0.31  & --0.21 &   9 & 0.17  & 0.14  & 110   \\ 
Reggiani et al.$^{11}$        & DEPM$^{12}$               &      6  &    0.00  & --0.02 &  14 &  -0.03  & 0.05  &  23   \\

        \hline
        \end{tabular}
        \\
$^1${$\Delta${\teff} = {\teff}(Author) -- {\teff}(Melendez)}, $^2${$\Delta$[Fe/H]  = [Fe/H](Author) -- [Fe/H](Melendez)}, $^3${$\Delta${\ali}  = {\ali}(Author) -- {\ali}(Melendez)}, $^4${Number of stars}, $^5${Slope of linear fit in Figure~\ref{fig:authors2}}, $^6${RMS about linear fit in Figure~\ref{fig:authors2}}, $^7${\citet{ryan99, ryan01b}}, $^8${Excitation Potential Method}, $^9${\citet{aoki09, aoki12, matsuno17a, matsuno17b, li15}},$^{10}${\citet{bonifacio12,bonifacio15,bonifacio18} (see Section~\ref{sec:abinitio})}, $^{11}${\citet{placco16}, \citet{reggiani17}}, $^{12}${Differential Excitation Potential Method}.\\

\end{table*}

\begin{figure*}
    \includegraphics[width=.80\hsize]{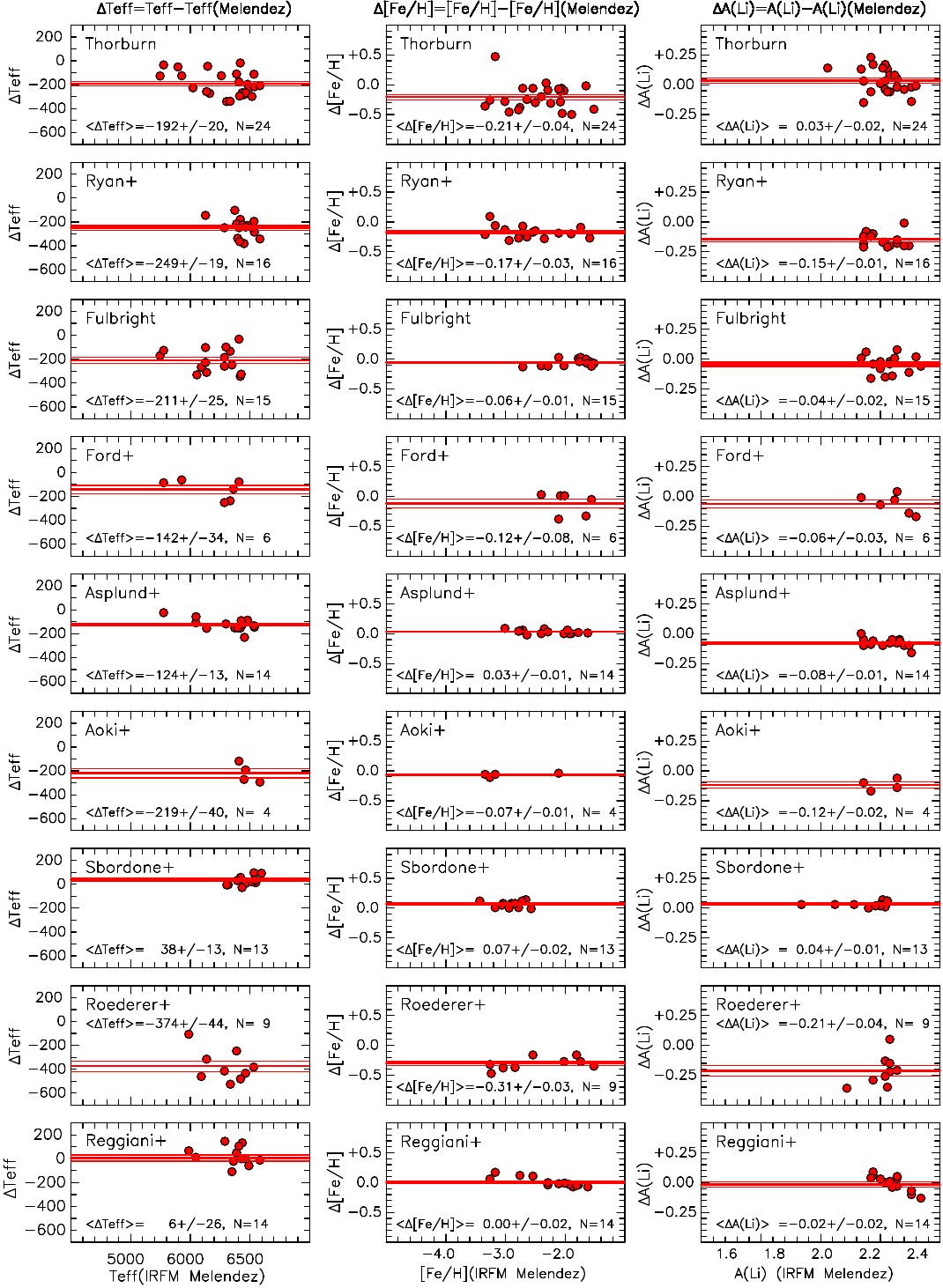}
    \caption{\small {$\Delta${\teff} vs. {\teff}(IRFM Melendez),
          $\Delta$[Fe/H] vs. [Fe/H](IRFM Melendez) and $\Delta${\ali}
          vs. {\ali}(IRFM Melendez)) for nine data subsets.  The
          authors, mean differences with errors, and number of data points appear
          within each panel.}}
    \label{fig:dteff}
\end{figure*}

\subsubsection{\it Systemic Differences Between Literature Values}

Perhaps the most important differences in Figure~\ref{fig:authors1} are those caused by systemic {\teff} differences.  In particular, the hotter stars of \citet{thorburn94}, \citet{ryan99, ryan01a, ryan01b}, \citet{fulbright00}, \citet{ford02}, \citet{asplund06}, Aoki and co-workers \citep{aoki09, aoki12, matsuno17a, matsuno17b, li15}, and \citet{roederer14} are some 200~K $-$ 350~K cooler than those of \citet{sbordone10} and \citet{melendez10}, both of which have IRFM-based {\teff}.  This is driven, in very large part, by significant differences in the assumptions and methods with which the temperatures have been derived.  We shall address this issue in Section~\ref{sec:corrected}.

We remind the reader that, fortuitously, {\ali}$_{\rm 1D,LTE}$ abundances differ only slightly from the more realistic {\ali}$_{\rm 3D,NLTE}$ values, with {\ali}$_{\rm 3D,NLTE}$ -- {\ali}$_{\rm 1D,LTE}$ $\sim$ 0.05, as first reported by \citet{asplund03} and \citet{asplund05}.

\subsubsection{\it What is the Lower {\teff} Limit of the Spite Plateau?} \label{sec:melendez10}
 
As noted by \citet{spite82}, the plateau effect in the {\ali} vs. {\teff} plane falls away at lower temperatures as a result of the increasing depth of the surface convective zone, which leads to lithium destruction by proton fusion.  The data of their Tables 3 and 5 define the Spite Plateau to lie within the ranges 5500~K $<$ {\teff} $<$ 6250~K and --2.4 $<$ [Fe/H] $<$ --1.1 (for objects on their {\teff} and [Fe/H] scales).  In the {\ali} vs. {\teff} panels in Figure~\ref{fig:authors1} we have therefore chosen to plot stars with {\teff} $>$ 5500~K.  It is clear, however, that in some half of the data sets in Figure~\ref{fig:authors1}, {\teff} = 5800~K would be a safer estimate of the limit (presumably as the result of different {\teff} scales).  Accordingly, in the {\ali} vs. [Fe/H] panels in Figure~\ref{fig:authors1} we have plotted only stars with {\teff} $>$ 5800~K.

\begin{figure*}
    \includegraphics[width=.85\hsize]{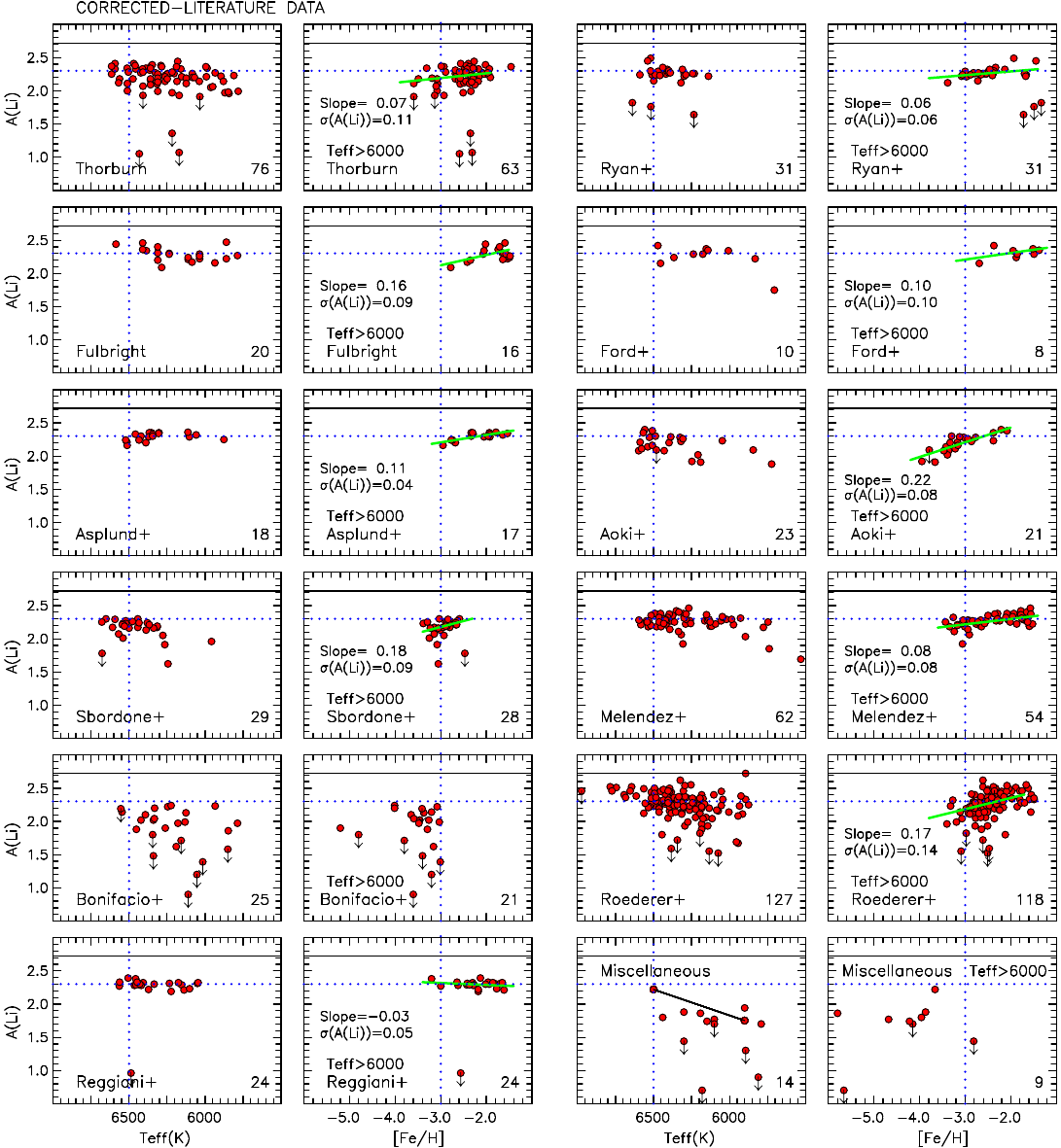}
    \caption{\small {{\ali} vs. {\teff} and [Fe/H] for the 12
          data samples, when the ``Literature'' values have been
          ``Corrected'' to the IRFM temperature scale of
          \citet{melendez10}. The format is the same as that of
          Figure~\ref{fig:authors1}.  See text for discussion.}}
    \label{fig:authors2}
\end{figure*}

\citet[see their Figure 2]{melendez10} report that the lower {\teff} limit of the plateau is a function of [Fe/H], increasing as [Fe/H] decreases.  The effect is evident in the Melendez et al. panel in our Figure~\ref{fig:authors1}, which is suggestive of a limit of {\teff} = 6000~K.  Further support for the concept is suggested by the components of the extremely-metal-poor (EMP), double-lined, spectroscopic binary CS~22876-032 which have {\teff} = 6500~K and 5900~K, [Fe/H] = --3.66 and --3.57, and the very different lithium abundances {\ali} = 2.22 and 1.75, respectively \citep{gonzalez08}. This effect can be seen in the bottom rightmost ({\ali}, {\teff}) panels of our Figures~\ref{fig:authors1} (and \ref{fig:authors2}), where the two stars are co-joined.  The simplest explanation of the large difference in lithium abundance of $\Delta${\ali} = 0.47 in this system is that not only does the effective temperature of the cooler secondary star lie below that of the cool edge of the Spite Plateau, but also that the cool edge of the plateau may have moved to even higher temperature for abundances below [Fe/H] = --3.6.

\begin{table*}
	\centering
	\caption{Multiple Observations of ``Literature'' and ``Corrected-Literature'' Data}
	\label{tab:tab3}
	\begin{tabular}{rlcrrrrrrrl}
	\hline
    Star No.  & 
	Starname &
	Coordinates (2000) & 
	{\teff} &
	{\logg} &
        [Fe/H] &
        {\ali} &
	{\teff} &
        [Fe/H] &
        {\ali} &
        Author$^1$ \\  
	    {} &	{} &	{} &	{(Lit)} &	{(Lit)} &	{(Lit)}&	{(Lit} &	{(Corr)} &	{(Corr)} &	{(Corr)}  & {}	\\
		{(1)} &	{(2)} &	{(3)} &	{(4)} &	{(5)} &	{(6)}&	{(7)} &	{(8)} &	{(9)} &	{(10)}  & {(11)}	\\\hline
  1  &  CS 22876-032A   &     00 07 37.10 --35 31 15.0 &  6319 &   4.00 &   --3.80 &     2.14 &    6511 &   --3.59 &     2.11 &  TH94   \\            
     &  CS 22876-032A   &     00 07 37.10 --35 31 15.0 &  6500 &   4.40 &   --3.66 &     2.22 &    6500 &   --3.66 &     2.22 &  GO08   \\            
  2  &  CS 22876-032B   &     00 07 37.10 --35 31 15.0 &  5900 &   4.60 &   --3.57 &     1.75 &    5900 &   --3.57 &     1.75 &  GO08   \\            
  3  &  LP 824-188      &     00 11 17.63 --20 43 30.7 &  5890 &   4.00 &   --1.84 &     2.07 &    6139 &   --1.67 &     2.22 &  RY99   \\            
  4  &  BS 17570-063    &     00 20 36.19  +23 47 37.6 &  6315 &   4.70 &   --2.86 &     2.09 &    6277 &   --2.93 &     2.05 &  SB10   \\            
     &  BS 17570-063    &     00 20 36.19  +23 47 37.6 &  6318 &   4.69 &   --2.91 &     2.06 &    6318 &   --2.91 &     2.06 &  ME10   \\            
  5  &  SD 0021--0050   &     00 21 13.78 --00 50 05.2 &  6546 &   4.30 &   --3.20 &     2.14 &    6750 &   --3.00 &     2.30 &  BO12   \\            
  6  &  SD 0023+0307    &     00 23 14.00  +03 07 58.1 &  6192 &   4.70 & $<-$5.80 &     1.86 &    6192 & $<-$5.80 &     1.86 &  AGFR   \\            
  7  &  SD 0027+1404    &     00 27 49.46  +14 04 18.1 &  6125 &   3.61 &   --3.37 &     2.13 &    6329 &   --3.17 &     2.29 &  BO12   \\            
  8  &  CS 29527-015    &     00 29 10.68 --19 10 07.3 &  6578 &   4.50 &   --3.31 &     2.27 &    6540 &   --3.38 &     2.23 &  SB10   \\            
     &  CS 29527-015    &     00 29 10.68 --19 10 07.3 &  6541 &   4.24 &   --3.43 &     2.25 &    6541 &   --3.43 &     2.25 &  ME10   \\            
  9  &  CS 30339-069    &     00 30 16.00 --35 56 55.0 &  6375 &   4.40 &   --2.98 &     2.20 &    6337 &   --3.05 &     2.16 &  SB10   \\            
 10  &  CS 22882-027    &     00 38 09.80 --31 47 58.0 &  6714 &   4.70 &   --2.40 &  $<$1.82 &    6676 &   --2.47 &  $<$1.78 &  SB10   \\            
 11  &  CD --33 239     &     00 39 51.92 --33 03 14.1 &  5993 &   4.00 &   --1.87 &     2.11 &    6242 &   --1.70 &     2.26 &  RY99   \\            
 12  &  SD 0040+1604    &     00 40 29.17  +16 04 16.2 &  6360 &   4.40 &   --3.29 &     1.99 &    6579 &   --3.22 &     2.11 &  AO09   \\            
     &  SD 0040+1604    &     00 40 29.17  +16 04 16.2 &  6422 &   3.90 &   --3.26 &     2.02 &    6626 &   --3.06 &     2.18 &  BO12   \\            
 13  &  BD +71 31       &     00 43 44.34  +72 10 43.1 &  6156 &   4.00 &   --2.23 &     2.42 &    6348 &   --2.02 &     2.39 &  TH94   \\            
 14  &  CS 22188-033    &     00 51 26.17 --38 12 17.8 &  6281 &   4.50 &   --2.98 &     1.66 &    6243 &   --3.05 &     1.62 &  SB10   \\            
 15  &  CS 29514-007    &     01 06 40.50 --24 58 41.0 &  6351 &   4.30 &   --2.79 &     2.23 &    6313 &   --2.86 &     2.19 &  SB10   \\            
 16  &  CS 29518-020    &     01 12 13.21 --31 00 05.3 &  6471 &   4.90 &   --2.60 &     2.28 &    6433 &   --2.67 &     2.24 &  SB10   \\            
     &  CS 29518-020    &     01 12 13.21 --31 00 05.3 &  6464 &   4.67 &   --2.72 &     2.25 &    6464 &   --2.72 &     2.25 &  ME10   \\            
 17  &  CS 29518-043    &     01 18 38.30 --30 41 02.8 &  6537 &   4.25 &   --3.16 &     2.20 &    6499 &   --3.23 &     2.16 &  SB10   \\            
     &  CS 29518-043    &     01 18 38.30 --30 41 02.8 &  6517 &   4.28 &   --3.17 &     2.20 &    6517 &   --3.17 &     2.20 &  ME10   \\            
 18  &  SD 0120--1001   &     01 20 32.63 --10 01 06.5 &  5627 &   3.40 &   --3.84 &     1.97 &    5846 &   --3.77 &     2.09 &  MA17   \\            
 19  &  CS 22953-037    &     01 25 06.80 --59 15 57.0 &  6557 &   4.45 &   --2.76 &     2.28 &    6519 &   --2.83 &     2.24 &  SB10   \\            
     &  CS 22953-037    &     01 25 06.80 --59 15 57.0 &  6532 &   4.33 &   --2.84 &     2.27 &    6532 &   --2.84 &     2.27 &  ME10   \\            
 20  &  SD 0140+2344    &     01 40 36.22  +23 44 58.0 &  5848 &   4.00 &   --4.00 &     1.86 &    6052 &   --3.80 &     2.02 &  BO18   \\            
 21  &  BD +02 263      &     01 45 13.81  +03 30 49.2 &  5800 &   4.00 &   --2.35 &     2.24 &    5992 &   --2.14 &     2.21 &  TH94   \\            
 22  &  G 245-32        &     01 47 12.38  +73 28 27.1 &  6290 &   4.00 &   --1.62 &     2.30 &    6539 &   --1.45 &     2.45 &  RY99   \\            
 23  &  BD --10 388     &     01 50 32.64 --09 21 02.8 &  6135 &   4.00 &   --2.29 &     2.30 &    6327 &   --2.08 &     2.27 &  TH94   \\            
     &  G 271-162       &     01 50 32.64 --09 21 02.8 &  6230 &   3.93 &   --2.30 &     2.27 &    6354 &   --2.33 &     2.35 &  AS06   \\            
     &  BD --10 388     &     01 50 32.64 --09 21 02.8 &  6260 &   3.98 &   --2.32 &     2.27 &    6260 &   --2.32 &     2.27 &  ME10   \\            
        
 	\hline
	\end{tabular}
	\\ 
	$^1$ Author: AGFR = \citep{aguado19, frebel19}, AO09 = \citet{aoki09}, AO12 = \citet{aoki12}, AS06 = \citet{asplund06}, BO12 = \citet{bonifacio12}, BO15 = \citet{bonifacio15}, BO18 = \citet{bonifacio18}, CA12 = \citet{caffau12}, CA16 = \citet{caffau16}, FO02 = \citet{ford02}, FR08 = \citet{frebel08}, FU00 = \citet{fulbright00}, GO08 = \citet{gonzalez08}, HA15 = \citet{hansen15}, LA21 = \citet{lardo21}, LI15 = \citet{li15}, MA17 = \citep{matsuno17a, matsuno17b}, ME10 = \citet{melendez10}, RE17 = \citep{placco16, reggiani17}, RY99 = \citep{ryan99, ryan01b}, SB10 = \citet{sbordone10}. SP13 = \citet{spite13}, ST18 = \citet{starkenburg18}, TH94 = \citep{thorburn94}
 \newline
	This table is published in its entirety in the electronic edition of the paper. A portion is shown here for guidance regarding its form and content.

\end{table*}
\newpage   

\begin{table*}
	\caption{{Averaged ``Literature'', ``Corrected-Literature'', and ``{\it ab initio} IRFM {\teff}'' Data}}
	\label{tab:tab4}
	\begin{tabular}{lccrrrrrrrrrr}
	\hline
        Starname     &
        Coordinates (2000)  &
        {\teff}      &
        [Fe/H]       &
        {\ali}       &
        {\teff}      &
        [Fe/H]       &
        {\ali}       &
        N            &
        {\teff}      &
        [Fe/H]       &
        {\ali}       \\
        {} &	{} &	{(Lit)} &	{(Lit)} &	{(Lit)} &	{(Corr)}&	{(Corr)} &	{(Corr)} &	{} &	{(IRFM)} & {(IRFM)} & {(IRFM)}\\
        
		{(1)} &	{(2)} &	{(3)} &	{(4)} &	{(5)} &	{(6)}&	{(7)} &	{(8)} &	{(9)} &	{(10)} & {(11)} & {(12)} \\
	\hline
CS 22876-032A   &     00 07 37.10 --35 31 15.0 &  6410 &   --3.73 &     2.18 &  6506 &   --3.62 &     2.16 &  2 &  6482 &   --3.62 &     2.09  \\     
CS 22876-032B   &     00 07 37.10 --35 31 15.0 &  5900 &   --3.57 &     1.75 &  5900 &   --3.57 &     1.75 &  1 &   ... &      ... &      ...  \\     
LP 824-188      &     00 11 17.63 --20 43 30.7 &  5890 &   --1.84 &     2.07 &  6139 &   --1.67 &     2.22 &  1 &  6098 &   --1.67 &     2.17  \\     
BS 17570-063    &     00 20 36.19  +23 47 37.6 &  6317 &   --2.88 &     2.07 &  6298 &   --2.92 &     2.05 &  2 &  5854 &   --2.92 &     1.69  \\     
SD 0021--0050   &     00 21 13.78 --00 50 05.2 &  6546 &   --3.20 &     2.14 &  6750 &   --3.00 &     2.30 &  1 &  6549 &   --3.00 &     2.14  \\     
SD 0023+0307    &     00 23 14.00  +03 07 58.1 &  6192 & $<-$5.80 &     1.86 &  6192 & $<-$5.80 &     1.86 &  1 &   ... &      ... &      ...  \\     
SD 0027+1404    &     00 27 49.46  +14 04 18.1 &  6125 &   --3.37 &     2.13 &  6329 &   --3.17 &     2.29 &  1 &  6235 &   --3.17 &     2.17  \\     
CS 29527-015    &     00 29 10.68 --19 10 07.3 &  6560 &   --3.37 &     2.26 &  6541 &   --3.41 &     2.24 &  2 &   ... &      ... &      ...  \\     
CS 30339-069    &     00 30 16.00 --35 56 55.0 &  6375 &   --2.98 &     2.20 &  6337 &   --3.05 &     2.16 &  1 &   ... &      ... &      ...  \\     
CS 22882-027    &     00 38 09.80 --31 47 58.0 &  6714 &   --2.40 &  $<$1.82 &  6676 &   --2.47 &  $<$1.78 &  1 &   ... &      ... &      ...  \\     
CD --33 239     &     00 39 51.92 --33 03 14.1 &  5993 &   --1.87 &     2.11 &  6242 &   --1.70 &     2.26 &  1 &  6413 &   --1.70 &     2.37  \\     
SD 0040+1604    &     00 40 29.17  +16 04 16.2 &  6391 &   --3.28 &     2.01 &  6603 &   --3.14 &     2.14 &  2 &  6275 &   --3.14 &     1.90  \\     
BD +71 31       &     00 43 44.34  +72 10 43.1 &  6156 &   --2.23 &     2.42 &  6348 &   --2.02 &     2.39 &  1 &   ... &      ... &      ...  \\     
CS 22188-033    &     00 51 26.17 --38 12 17.8 &  6281 &   --2.98 &     1.66 &  6243 &   --3.05 &     1.62 &  1 &  6365 &   --3.05 &     1.72  \\     
CS 29514-007    &     01 06 40.50 --24 58 41.0 &  6351 &   --2.79 &     2.23 &  6313 &   --2.86 &     2.19 &  1 &  6357 &   --2.86 &     2.22  \\     
CS 29518-020    &     01 12 13.21 --31 00 05.3 &  6468 &   --2.66 &     2.26 &  6449 &   --2.70 &     2.24 &  2 &  6501 &   --2.70 &     2.27  \\     
CS 29518-043    &     01 18 38.30 --30 41 02.8 &  6527 &   --3.16 &     2.20 &  6508 &   --3.20 &     2.18 &  2 &  6576 &   --3.20 &     2.22  \\     
SD 0120--1001   &     01 20 32.63 --10 01 06.5 &  5627 &   --3.84 &     1.97 &  5846 &   --3.77 &     2.09 &  1 &   ... &      ... &      ...  \\     
CS 22953-037    &     01 25 06.80 --59 15 57.0 &  6545 &   --2.80 &     2.28 &  6526 &   --2.84 &     2.26 &  2 &  6684 &   --2.84 &     2.35  \\     
SD 0140+2344    &     01 40 36.22  +23 44 58.0 &  5848 &   --4.00 &     1.86 &  6052 &   --3.80 &     2.02 &  1 &  5582 &   --3.80 &     1.56  \\     
BD +02 263      &     01 45 13.81  +03 30 49.2 &  5800 &   --2.35 &     2.24 &  5992 &   --2.14 &     2.21 &  1 &  6111 &   --2.14 &     2.26  \\     
G 245-32        &     01 47 12.38  +73 28 27.1 &  6290 &   --1.62 &     2.30 &  6539 &   --1.45 &     2.45 &  1 &   ... &      ... &      ...  \\     
BD --10 388     &     01 50 32.64 --09 21 02.8 &  6208 &   --2.30 &     2.28 &  6314 &   --2.24 &     2.30 &  3 &  6220 &   --2.24 &     2.17  \\     
\hline
	\end{tabular}
	\\
	This table is published in its entirety in the electronic edition of the paper. A portion is shown here for guidance regarding its form and content.
\end{table*}
\newpage   

\subsection{Step 2 -- ``Corrected'' Effective Temperatures on the IRFM Scale} \label{sec:corrected}

{\ali} is sensitive to the differences in {\teff} resulting from the several methods adopted in the literature.  As noted above, systemic differences in the manner in which literature {\teff} values are determined cause differences of order 200~K $-$ 350~K.  An error of 300~K results in the lithium abundance changing by $\Delta${\ali} $\sim$~0.25 dex.  To address this issue we sought to ``correct'' all literature {\teff} values to the IRFM scale, by adopting the IRFM {\teff} values of \citet{melendez10} as standards, and which we shall refer to as ``Corrected-Literature'' values. We regard the IRFM method adopted by Melendez et at (2010), in particular the use of interstellar Na~D lines to determine {\ebv} values, as a positive step forward.

Figure~\ref{fig:dteff} and Table~\ref{tab:tab2} quantify the mean differences of $\Delta${\teff} = {\teff}(Author) -- {\teff}(Melendez) vs. {\teff}(Melendez), $\Delta$[Fe/H] = [Fe/H](Author) -- [Fe/H](Melendez) vs. [Fe/H](Melendez), and $\Delta${\ali} = {\ali}(Author) -- {\ali}(Melendez) vs. {\ali}(Melendez), for the nine literature subsets that have a sufficient overlap of stars with those of \citet{melendez10}.  These differences are driven essentially by those in {\teff} between the literature temperature scales and the IRFM scale adopted by \citet{melendez10}, and bring the literature values onto the IRFM scale.  Columns (1) and (2) of the table contain the authorship and methods used to determine {\teff}, while columns (3) $-$ (6) present the resulting $\langle\Delta${\teff}$\rangle$, $\langle\Delta$[Fe/H]$\rangle$, $\langle\Delta${\ali}$\rangle$, and number of stars, respectively.

The two literature subsets that do not have sufficient overlap with \citet{melendez10} are those of \citet{bonifacio12, bonifacio15, bonifacio18} and the twelfth subset of 14 miscellaneous stars having [Fe/H] $<$ --2.8 (occupying the bottom right panels of Figure~\ref{fig:authors1}).  In Section~\ref{sec:abinitio} we shall discuss IRFM corrections to the literature values of {\teff}, [Fe/H], and {\ali} of the Bonifacio et al. data, which we have also included in Table 2.  For the ``miscellaneous'' subset we have applied no correction.

Figure~\ref{fig:authors2}, which has the same format as Figure~\ref{fig:authors1}, presents the ``Corrected'' results when the differences in Table~\ref{tab:tab2} (including those of Bonifacio et al.) are applied to the literature {\teff}, [Fe/H], and {\ali} values.  In Figure~\ref{fig:authors1} we applied an exclusion of stars having {\teff} $<$ 5800~K from the ({\ali}, [Fe/H]) -- plane.  In view of the corrections that have been made to the literature {\teff} values, which are of order +200~K, we have adopted {\teff} $<$ 6000~K as being at or above the cooler limit of the Spite Plateau in the ({\ali}, [Fe/H]) -- plane of Figure~\ref{fig:authors2}.  The slopes and RMS dispersions of the data about the linear lines of best fit in the ({\ali}, [Fe/H]) -- planes (introduced in Figure~\ref{fig:authors1}) are presented in Figure~\ref{fig:authors2} and in columns (7) and (8) of Table~\ref{tab:tab2}.

A comparison between Figures~\ref{fig:authors1} and \ref{fig:authors2}
reveals the following: (1) there is better agreement in
Figure~\ref{fig:authors2} between the effective temperatures of the
hotter stars of a majority of the panels; (2) there are subtle
improvements in {\ali} values bringing them closer to a common
plateau; (3) a large dispersion of $\sigma$({\ali}) = 0.14 remains in
the \citet{roederer14} sample, which is not too surprising insofar as
they estimate that the error in their Li abundances is $\sim$ 0.10 --
0.20~dex; and (4) the Bonifacio et al. data, which have [Fe/H] $<$
--3.0, stretch to lower [Fe/H] than the other samples and show large
scatter.  This dispersion is the Sbordone et al. and Bonifacio et
al. ``meltdown''.

Given the larger size of the \citet{roederer14} errors relative to those of the other samples, we choose not to include this dataset further in our discussions of lithium abundances.

Table~\ref{tab:tab3} presents the ``Literature'' and ``Corrected-Literature'' of 332 literature observations of the 195 individual stars in the data sets presented in Figures~\ref{fig:authors1} and \ref{fig:authors2} (excluding the \citealt{roederer14} data).  Column (1) contains the star number, (2) and (3) the star name and coordinates, (4) $-$ (7) present {\teff}, {\logg}, [Fe/H], {\ali} for the ``Literature'' data set, (8) $-$ (10) have {\teff}, [Fe/H], and {\ali} for the ``Corrected-Literature'' sample, and (11) shows code for the original data sources.

\begin{figure}
   \includegraphics[width=.80\hsize]{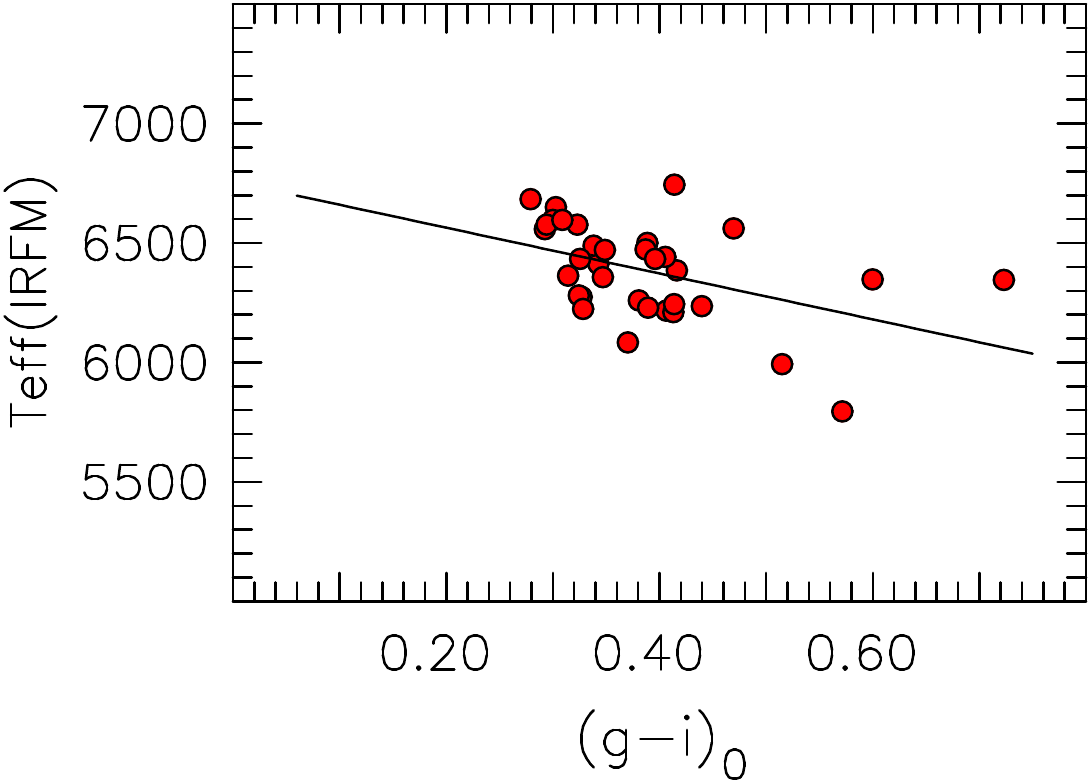}
   \caption{\small {{\tirfm} vs. {\gizero}: {\tirfm}({\gizero}) =
        6756.4 -- 959.7{\gizero}~(K) (34 stars, yielding RMS =
       182~K).}}
   \label{fig:grit196}
\end{figure}

\begin{figure*}
   \includegraphics[width=0.9\hsize]{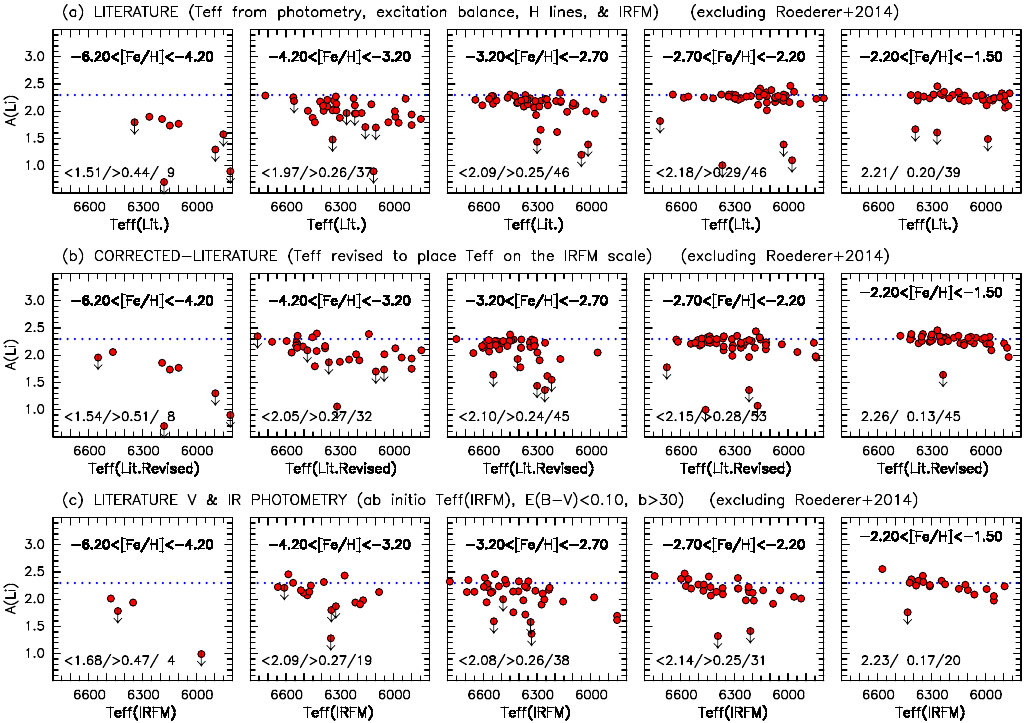}
   \caption{\small {{\ali} vs. {\teff} for (a) the ``Literature'',
         (b) ``Corrected-Literature'', and (c) ``{\it ab initio}
         {\tirfm}'' datasets (excluding the \citealt{roederer14}
         sample). In the five panels of each row the range of [Fe/H]
         is presented towards the top, with [Fe/H] increasing from
         left to right.  At the bottom left of each panel the
          three numbers represent the mean {\ali}, its dispersion $\sigma$,
         and number of stars.}}
   \label{fig:lastly196}
\end{figure*}

\begin{figure}
    \includegraphics[width=0.575\hsize]{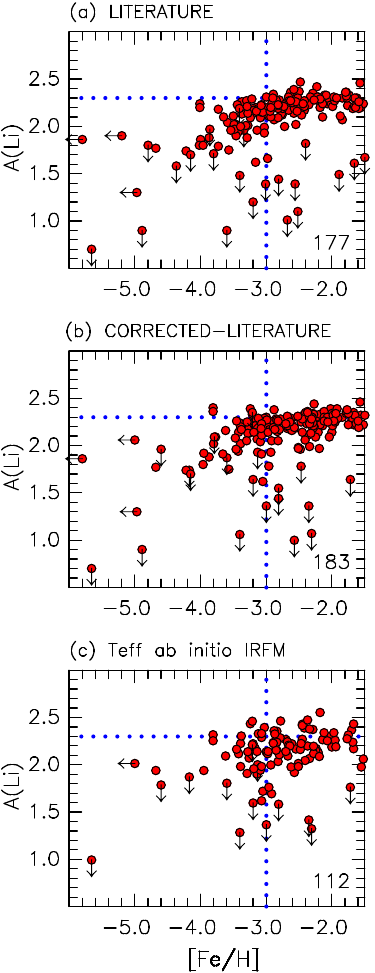}
    \caption{\small {{\ali} vs. [Fe/H] for the (a) ``Literature'', (b)
        ``Corrected-Literature'', and (c) ``{\it ab initio} {\tirfm}''
        datasets (excluding the \citealt{roederer14} sample). The number
        of stars is given at the bottom-right of each panel.}}
    \label{fig:lastly196alife}
\end{figure}

\begin{figure}
    \includegraphics[width=.75\hsize]{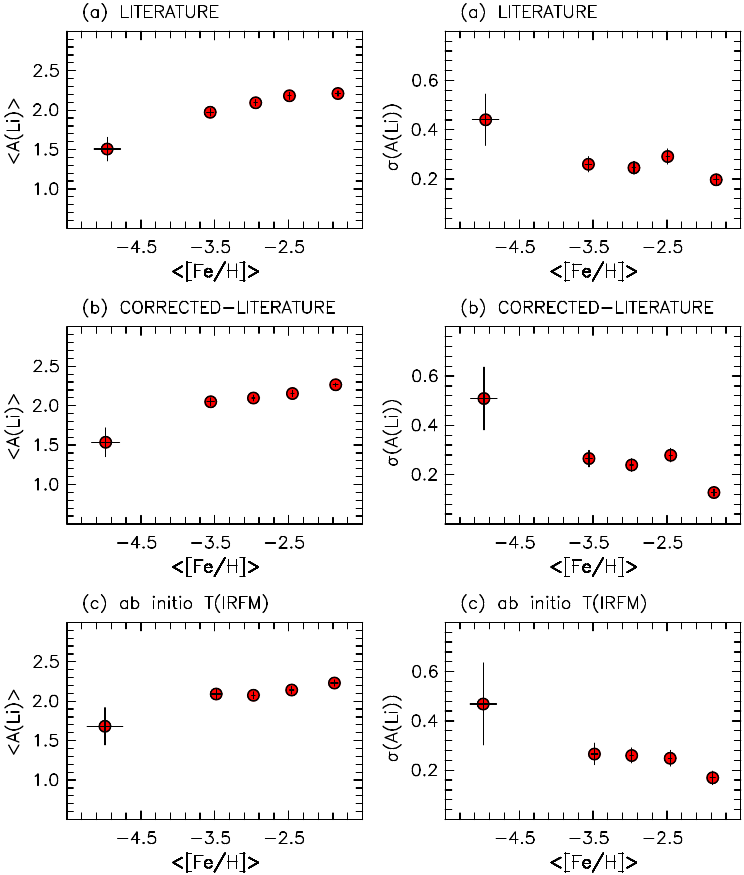}
    \caption{\small {Averages $\langle${\ali}$\rangle$ (left) and
          dispersions, $\sigma$({\ali}) (right)
          vs. $\langle$[Fe/H]$\rangle$ for stars in the five
          horizontal ``Corrected-Literature'' panels of
          Figure~\ref{fig:lastly196}.}}
   \label{fig:endplot196}
\end{figure}

\begin{figure}
    \includegraphics[width=.625\hsize]{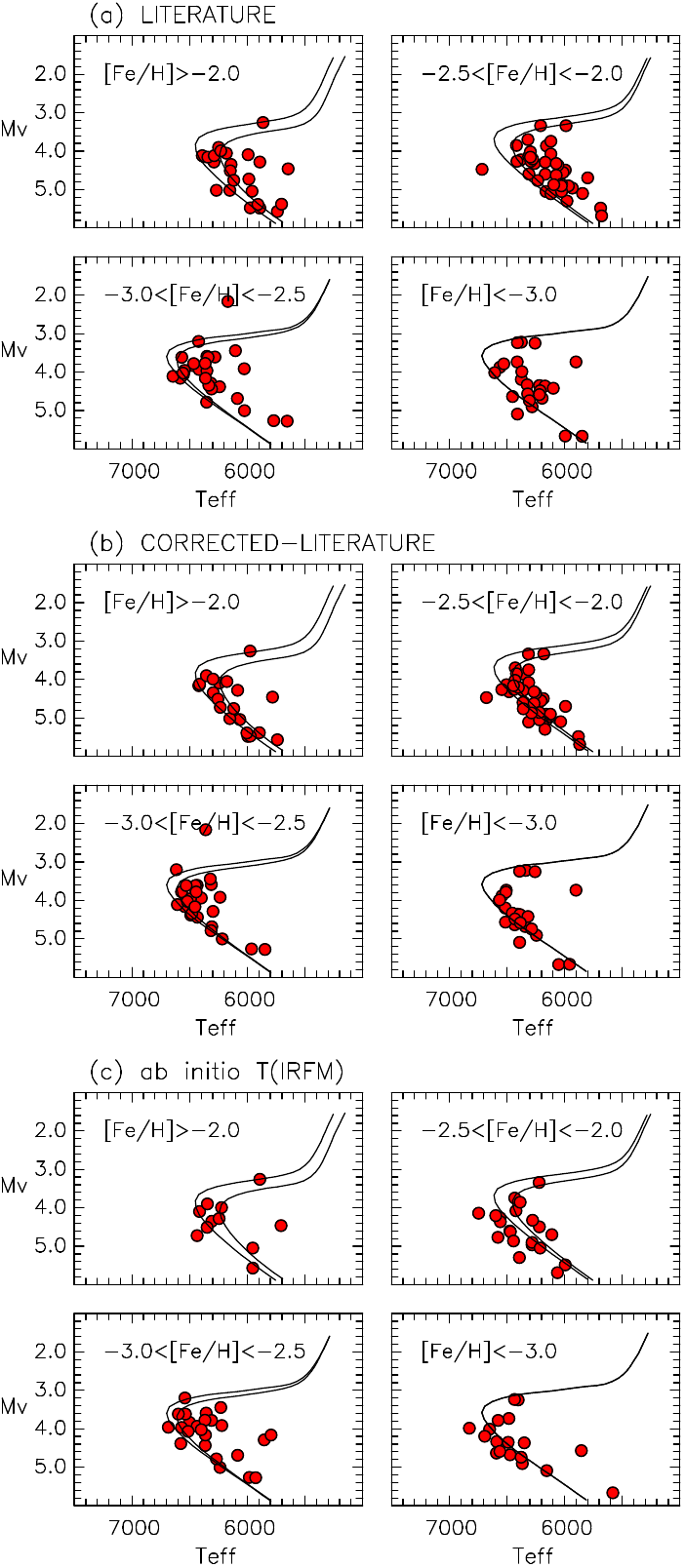}
    \caption{\small {({$M_{V}$}, {\teff}) --
    Colour-Magnitude Diagrams for (a) the ``Literature'', (b)
    ``Corrected-Literature'', and (c) ``{\it ab initio} {\tirfm}''
    data sets.  In each set the four sub-panels pertain to stars
    within the indicated [Fe/H] limits.  Also presented are
    Yale-Yonsei Isochrones for these [Fe/H] values, helium abundance Y
    = 0.23, [$\alpha$/Fe] = 0.3, and Age = 12 Gyr.}}
    \label{fig:cmd196}
\end{figure}

\subsection{Step 3 - {\bf {\it ab initio}} Temperatures on the IRFM Scale} \label{sec:abinitio}

To complement the two previous steps, we analyzed the literature equivalent widths of the Li I 6707~{\AA} doublet\footnote{For the some 40 stars in the literature data that have {\ali} but no published equivalent widths, we used the \citet[Table B.1]{sbordone10} algorithm, together with literature {\ali}, {\teff}, [Fe/H], and {\logg} to obtain those values.} together with independent {\it ab initio} IRFM {\teff} values based on visual and infrared colours, using the algorithm of \citet[Table B.1]{sbordone10} to determine {\ali}$_{\rm 1D,LTE}$ values.  This procedure also requires input [Fe/H] values, to which it is relatively insensitive\footnote{By way of example, at {\tirfm} = 6000~K, a change of $\Delta$[Fe/H] of 0.5~dex causes a change in {\tirfm} of order 20 $-$ 50~K.}, and for which we adopted the ``corrected'' [Fe/H] values of Step 2.  We used literature {\logg} values (or {\logg} = 4.0 in their absence), to which the process is insensitive for near-main-sequence stars.

In order to determine IRFM {\teff} (hereafter {\tirfm}) from visual and infrared observations ({\vi}, {\vj}, and {\vk} data) we adopted photometric data from the Simbad archives, and UCAC4 \citep{zacharias12} and APASS \citep{henden16} values from the VizieR archives{\footnote{https://simbad.unistra.fr/simbad/ and https://vizier.u-strasbg.fr. Magnitudes were accepted only if errors were less than 0.03 for {\V}, {\I}, {\J}, and {\K} and 0.05 for {\g} and {\ii}.}.  Reddening estimates followed \citet[Section 4.1]{melendez06}\footnote{We are grateful to J. Melendez for making available the Fortran code ``extinct.for'' of \citet{hakkila97}.} and we accepted stars having {\ebv} $\leq$ 0.10 and Galactic latitude $|$b$|$ $>$ 30.  All reddenings determined in the present paper have utilized this procedure, using distances obtained from the Gaia EDR3 archive (https://gea.esac.esa.int/archive/); see also \citet{gaia16} and \citet{gaia21} and the IRFM ({\teff}, colour) fits of \citet{casagrande10}.

As foreshadowed above for the Bonicacio et al. stars, which do not have sufficient overlap with the \citet{melendez10} subset, we proceeded as follows.  Using stars in our sample that have both {\tirfm} and APASS {\gi} values, we determined the linear least-squares fit to {\tirfm} vs. {\gizero}: {\tirfm}({\gizero}) = 6756.4 -- 959.7{\gizero}~(K) (34 stars, yielding RMS = 182~K). The fit to the data is presented in Figure~\ref{fig:grit196}.  We then used the SDSS {\gi} values from \citet{bonifacio12, bonifacio15, bonifacio18} to obtain temperatures using this relationship.  It is of interest to compare our {\tirfm}({\gizero}) values with the infrared temperatures of Bonifacio et al.  While their {\teff} values were based on SDSS ($g$--$z$) colours, ours utilize SDSS ($g$--$i$).  From 25 stars in common, we find a mean difference $\langle$$\Delta${\teff}$\rangle$ = $\langle${\teff}(Bonifacio et al.) -- {\tirfm}({\gizero})$\rangle$ = --204 $\pm$ 21~K.  We adopt this result to determine IRFM corrections to the Bonifacio et al. literature values in Table~\ref{tab:tab2}: $\Delta${\teff} = --204~K, $\Delta$[Fe/H] = --0.20 ({\it cf.} \citealt{yong13a}) and $\Delta${\ali} = --0.16, (utilizing the \citet[Table B.1]{sbordone10} algorithm). 

Table~\ref{tab:tab4} presents the averaged data for the 195 individual stars.  Columns (1) and (2) contain star name and coordinates, (3) $-$ (5) present {\teff}, [Fe/H], {\ali} for the ``Literature'' data set, while (6) $-$ (8) have the same parameters for the ``Corrected-Literature'' sample, respectively.  Column (9) contains the number of data sources included in the average for each star, while Columns (10) $-$ (12) list {\teff}, [Fe/H], and {\ali} values for 121 stars from the ``{\it ab initio} IRFM" step.

\subsection{Comparison of the Three Data Steps} \label{sec:3steps}

At this point we have up to three independent {\ali} values for each of the 195 individual stars. Figure~\ref{fig:lastly196} presents the dependence of {\ali} vs. {\teff} as a function of [Fe/H] for the above three steps, where rows (a) $-$ (c) contain the ``Literature'', ``Corrected-Literature'', and ``{\it ab initio} IRFM'' results, respectively.  In each row of the figure there are five panels containing ({\ali}, {\teff}) values for stars in the [Fe/H] range indicated within each panel.  Note that the [Fe/H] ranges are the same in each column of the figure.  In all panels the blue horizontal dotted line represents {\ali} = 2.3, which is useful to facilitate comparison of the results.  At the bottom left of each panel the three numbers represent the mean {\ali}, its dispersion $\sigma$({\ali}) , and the number of stars.  For completeness, in Figure~\ref{fig:lastly196alife} we show the complementary {\ali} vs. [Fe/H] diagrams for the three cases.

While the three rows in Figure~\ref{fig:lastly196} show slight differences, there are clear similarities. Here we shall discuss our conclusions with reference to the ``Corrected-Literature'' middle row, which we regard as the most reliable of the three.  Recalling that the {\ali} vs. {\teff} format was the one first used by \citet{spite82} to define what became known as the Spite Plateau, we make the following points: (1) in the rightmost panel, which covers --2.2 $<$ [Fe/H] $<$ --1.5, the lithium abundances are constant for {\teff} $>$ 6000~K, as reported by \citet{spite82}; then (2) as one proceeds from higher to lower [Fe/H] values in the figure, $\langle${\ali}$\rangle$ decreases; while (3) the dispersion $\sigma$({\ali}) increases.  In the bottom of each panel the three numbers are $\langle${\ali}$\rangle$, $\sigma${(\ali}), and number of stars, respectively.  These data are also presented, together with $\langle$[Fe/H]$\rangle$, in Table~\ref{tab:tab5}, and plotted in Figure~\ref{fig:endplot196}, where the left column contains the dependence of $\langle${\ali}$\rangle$ on $\langle$[Fe/H]$\rangle$, and the right presents the dispersion, $\sigma$({\ali}) vs. $\langle$[Fe/H]$\rangle$.

\subsubsection{Colour-Magnitude Diagrams (CMDs) for the Three Data Steps}

A byproduct of the present investigation is that given {\teff}, $V$ magnitudes, and distances one is able to construct ({$M_{V}$}, {\teff})   $-$ colour-magnitude diagrams, and to compare them with theoretical isochrones in an effort to test the {\teff} values determined above.  In Figure~\ref{fig:cmd196} we present {$M_{V}$} vs. {\teff} CMDs for the three {\teff} cases discussed above, where each of the three groupings in the figure contains four sub-panels embracing different [Fe/H] ranges, together with the Yale-Yonsei isochrones of \citet{demarque04}\footnote{http://www.astro.yale.edu/demarque/yyiso.html} adopting an age of 12 Gyr, for the ranges presented in the sub-panels.  Inspection of the middle panel, which contains the ``Corrected-Literature'' results, offers strong and reassuring support for the procedures we have adopted in Section~\ref{sec:corrected}.

\section{Carbon Abundances, [C/Fe], in the Range --6.0 $<$ [Fe/H] $<$ --2.0}

We noted in our Table~\ref{tab:tab1} that \citet{matsuno17b} found no apparent anticorrelation between lithium and carbon at [Fe/H] $\sim$ --3.0 $-$ specifically, they reported the existence of two C-rich (CEMP-no) stars ``near the plateau'' with lithium abundances {\ali} = 2.17 and 2.14.  \citet{jacobson15} and \citet{placco16} had previously presented similar carbon abundances for three CEMP-no stars which have {\ali} values in the range 1.99 $-$ 2.36.  For completeness, the [Fe/H], [C/Fe], [Ba/Fe], and {\ali} values for these five stars are presented in Table~\ref{tab:tab6}.

\begin{table}
	\caption{Average ``Corrected-Literature'' {\ali} and [Fe/H] Data from Figure~6}
	\label{tab:tab5}
	\begin{tabular}{cccc}
	\hline
	$\langle${\ali}$\rangle$ &
        $\sigma$({\ali})         &
	$\langle$[Fe/H]$\rangle$ &
	N$^1$                         \\
	{(1)} &	{(2)} &	{(3)} &	{(4)} \\
	\hline
        1.536   &   0.509   &  --4.977   &     8  \\
        2.049   &   0.265   &  --3.556   &    32  \\
        2.097   &   0.239   &  --2.977   &    45  \\
        2.155   &   0.279   &  --2.448   &    53  \\
        2.264   &   0.127   &  --1.864   &    45  \\
	\hline
	\end{tabular}

        $^1${Number of stars}
\end{table}
 
q\begin{table*}
	\caption{Five CEMP-no stars near [Fe/H] $\sim$ --3.0}
	\label{tab:tab6}
	\begin{tabular}{lccrcc}
	\hline
	Starname &
	[Fe/H]   &
        [C/Fe]   &
	[Ba/Fe]  &
	{\ali}   &
        Author$^1$  \\
	{(1)} &	{(2)} &	{(3)} &	{(4)} &	{(5)} &	{(6)} \\
	\hline
	CD --24 17504       &  --3.41  &   1.10 &  $<$--1.05 &  1.99  &   1    \\
	G 64-12             &  --3.29  &   1.07 &     --0.07 &  2.36  &   2    \\
	G 64-37             &  --3.11  &   1.12 &     --0.36 &  2.25  &   2    \\
	LA 1410--0555$^2$   &  --3.19  &   1.53 &     --0.33 &  2.17  &   3    \\
	SD 1424+5615$^2$    &  --3.01  &   1.49 &  $<$--0.69 &  2.14  &   3    \\
	\hline
	\end{tabular} \\
        $^1$1 = \citet{jacobson15}, 2  = \citet{placco16}, 3  = \citet{matsuno17b}
        \newline $^2${LA 1410--0555 = {LAMOST J1410--0555}, SD 1424+5615 = SDSS J1424+5615}
\end{table*}

\begin{table*}
	\caption{[C/Fe] for 78 Near-Main-Sequence Stars over Three Abundance Ranges}
	\label{tab:tab7}
	\begin{tabular}{lccrrl}
	\hline
    Starname &
    {\teff}  &
	[Fe/H]   &
	{\ali}   &
    [C/Fe]   &
    Author$^1$  \\
	{(1)} &	{(2)} &	{(3)} &	{(4)} &	{(5)} &	{(6)} \\
	\hline
\multicolumn{6}{c}{[Fe/H] $<$ $-$4.5} \\
SD 0023+0307     &  6192  & $<$--5.80  &    1.86  & $>$3.77  &  AGFR     \\     
HE 0233--0343    &  6100  &    --4.68  &    1.77  &    3.48  &  HA15     \\     
SD 0929+0238     &  5894  & $<$--4.97  & $<$1.30  & $>$4.24  &  CA16     \\     
SD 1029+1729     &  5811  &    --4.89  & $<$0.90  & $<$1.00  &  CA12     \\     
SD 1035+0641     &  6466  & $<$--5.00  &    2.06  & $>$3.73  &  BO18     \\     
HE 1327--2326    &  6180  &    --5.66  & $<$0.70  &    4.13  &  FR08     \\     
Pr 221.8+09.7     &  5792  &    --4.66  &    1.70  & $<$1.83  &  ST18     \\     
SD 1742+2531     &  6549  &    --4.60  & $<$1.96  &    3.43  &  BO15     \\     
\multicolumn{6}{c}{$-$4.2 $<$ [Fe/H] $\leq$ $-$3.0}  \\
CS 29527-015     &  6541  &    --3.41  &    2.24  &    1.13  &  BO09     \\     
CS 30339-069     &  6337  &    --3.05  &    2.16  &    0.62  &  BO09     \\     
CS 22188-033     &  6243  &    --3.05  &    1.62  & $<$0.84  &  MP21     \\     
SD 0120--1001    &  5846  &    --3.77  &    2.09  & $<$1.81  &  MA17     \\     
SD 0140+2344     &  6052  &    --3.80  &    2.02  &    2.00  &  BO18     \\     
HE 0148--2611    &  6568  &    --3.14  &    2.07  & $<$1.16  &  MP21     \\     
CS 22958-042     &  6409  &    --3.13  & $<$1.93  &    3.15  &  SI06     \\     
SD 0212+0137     &  6537  &    --3.39  &    2.20  &    2.08  &  BO15     \\     
\multicolumn{6}{c}{$-$3.0 $<$ [Fe/H] $\leq$ $-$2.0}  \\
BS 17570-063     &  6298  &    --2.92  &    2.05  &    0.49  &  BO09     \\     
CS 22882-027     &  6676  &    --2.47  & $<$1.78  & $<$0.24  &  MP21     \\     
CS 29514-007     &  6313  &    --2.86  &    2.19  & $<$0.76  &  MP21     \\     
CS 22953-037     &  6526  &    --2.84  &    2.26  &    0.41  &  BO09     \\     
CS 31061-032     &  6400  &    --2.61  &    2.22  &    0.68  &  BO09     \\     
LP 651-4         &  6489  &    --2.79  &    2.26  &    0.57  &  NO19     \\     
G 4-37           &  6308  &    --2.57  &    2.18  &    0.60  &  NO19     \\
HD 19445         &  6087  &    --2.01  &    2.27   &   0.45  &  TO92     \\     
	\hline
	\end{tabular}
	\\
$^1$AGFR = \citet{aguado19}, \citet{frebel19}, AL15 = \citet{alex15}, BA05 = \citet{barklem05}, BO09 = \citet{bonifacio09}, BO15 = \citet{bonifacio15}, BO18 = \citet{bonifacio18},
        CA11 = \citet{caffau11},
        CA16 = \citet{caffau16},
        FR08 = \citet{frebel08},
        HA15 = \citet{hansen15},
        JA15 = \citet{jacobson15},
        LA08 = \citet{lai08},                  
        LI15 = \citet{li15},                   
        MA17 = \citet{matsuno17a,matsuno17b},
        NO19 = \citet{norris19},               
        PL16 = \citet{placco16},
        ST18 = \citet{starkenburg18},
        SI06 = \citet{siva06},
        TO92 = \citet{tomkin92} \\
	This table is published in its entirety in the electronic edition of the paper. A portion is shown here for guidance regarding its form and content.
\end{table*}

\begin{figure*}
    \includegraphics[width=.70\hsize]{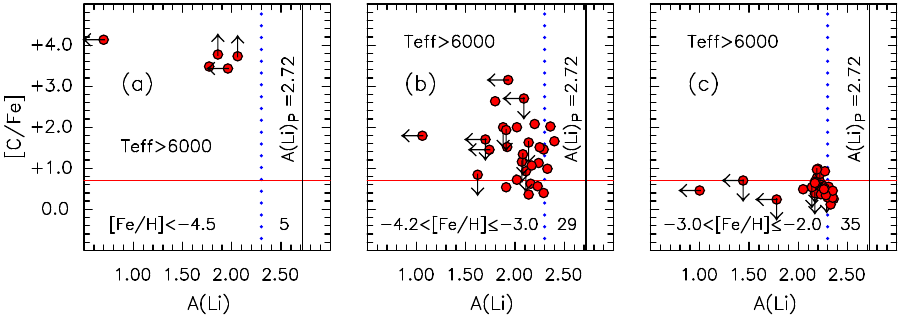}
    \caption{\small {[C/Fe] vs. {\ali} for stars with (a) [Fe/H] $<$
          --4.5, (b) --4.2 $<$ [Fe/H] $\leq$ --3.0, and (c) --3.0 $<$ [Fe/H] $\leq$ --2.0, for stars with {\teff} $>$ 6000~K. The number of stars is presented at the bottom right. Black vertical lines represent Primordial Lithium ({\ali}$_{\rm P}$ = 2.72), the red line at {\ali} = 0.7 separates C-weak and C-strong stars, and the blue dotted lines are for reference purposes.}}
    \label{fig:licfe_obs}
\end{figure*}

As discussed above, it is well-established that the large majority of stars with [Fe/H] $<$ --4.5 are C-rich, and that the majority of those near the main-sequence are also Li-poor.  We now investigate to what extent, if any, that relationship exists also in the range --4.2 $<$ [Fe/H] $\leq$ --2.0.  That is to say, is there any connection between {\ali} and [C/Fe] in or near the ``meltdown'' zone?

[C/Fe] values from the literature are presented for 78 near-main-sequence stars having [Fe/H] $\leq$ --2.0 in Table~\ref{tab:tab7}\footnote{We note that the [C/Fe] abundances in Table~\ref{tab:tab7} have been rescaled from the literature values to adopt the [Fe/H] of the table and a solar carbon abundance A(C)$_\odot$ = 8.43 \citep{asplund09}.}.  The data derive principally from the subsets discussed in Section~\ref{sec:critique}: \citet{frebel08, frebel19}, \citet{bonifacio09, bonifacio12, bonifacio18}, \citet{caffau11, caffau12, caffau16}, \citet{spite13}, \citet{hansen15}, \citet{jacobson15}, \citet{li15}, \citet{placco16}, \citet{matsuno17a, matsuno17b}, \citet{starkenburg18}, \citet{aguado19},and \citet{lardo21}, and are supplemented by material from \citet{tomkin92}, \citet{honda04}, \citet{barklem05}, \citet{siva06}, \citet{lai08}, \citet{alex15}, \citet{norris19}, and \citet{matas-pinto21}.  Figure~\ref{fig:licfe_obs} presents [C/Fe] vs. {\ali} for three [Fe/H] ranges, for stars having {\teff} $>$ 6000~K: (a) [Fe/H] $<$ --4.5 (the most-Fe-poor observed stars), (b) --4.2 $<$ [Fe/H] $\leq$ --3.0 (the ``meltdown'' region), and (c) --3.0 $<$ [Fe/H] $\leq$ --2.0 (our upper [Fe/H] region). 

As part of our investigation we sought to confirm the [C/Fe] abundances of four stars in Table~\ref{tab:tab7} that have both low {\ali} and low [C/Fe] and for which spectra are available in the ESO UVES archives, using the techniques adopted in \citet{norris19}. These are CS~22882-027 with ({\ali}, [C/Fe]) = ($<$1.78, $<$0.24), CS~22188-033 (1.62, $<$0.84), BS~17570-063 (2.05, 0.49), and CS~22966-011 (1.91, 0.54).  Our results were essentially the same as reported by the authors (\citealt{bonifacio09} and \citealt{matas-pinto21}) in Table~\ref{tab:tab7}, except that for BS~17570-063 and CS~22966-011 we obtained abundances that we regarded as limits rather than detections.

Defining C-rich stars as those with [C/Fe] $>$ 0.7 (following \citealt{aoki07}) one finds that the most metal-poor regime ([Fe/H] $<$ --4.5) in Table~\ref{tab:tab7} has an extremely high C-rich fraction of 1.00, as might be expected from our previous discussion. The potentially more significant and important result in the present context, however, appears in the ``meltdown'' regime (--4.2 $<$ [Fe/H] $\leq$ --3.0), where the incidence of stars with [C/Fe] $>$ 0.7 is very high. That is, the fraction of stars with detected carbon and [C/Fe] $>$ 0.7 is 0.52.  If any of the stars with a limit above [C/Fe] = 0.7 turned out to be C-rich, the C-rich fraction would be greater that 0.52.\footnote{We estimate that for a star with {\teff} = 6000~K, {\logg} = 4.5, and [O/Fe] = +0.4 to have a detectable CH 4323~{\AA} feature at the 0.98 intensity level relative to the continuum would require [C/Fe] = 1.3.}

Inspection of the data in Table ~\ref{tab:tab7} permits us to investigate more closely the stars in the range --3.0 $<$ [Fe/H] $<$ --2.5 that have the smallest [C/Fe] values. Of particular interest for the present discussion are the four stars in Figure~\ref{fig:licfe_obs} that have [C/Fe] $<$ 0.9 and {\ali} $<$ 1.8, and which offer no support for an anticorrelation of {\ali} and [C/Fe].  They are CS22882-027 with ({\ali}, [C/Fe]) = ($<$1.78, $<$ 0.24), CS22188-033 (1.62, $<$ 0.84), HE0411-3558 ($<$1.44, $<$ 0.70), and G186-26 ($<$1.00, 0.46).  Three of these have been identified as belonging to the ``blue-straggler'' related group discussed above $-$ the first two by \citet{matas-pinto21} and the fourth by \citet{ryan01a}, suggesting they are not relevant to the question at issue here.

We shall return, in Section~\ref{sec:toymod} in our discussion of the ``meltdown'', to a possible explanation of the data in Figure~\ref{fig:licfe_obs}.

\section{Discussion} \label{sec:discussion}

Given that the Li abundances in Figure~\ref{fig:lastly196} are strongly dependent on [Fe/H], which to first order increases with time in the early universe, we are of the view that an explanation of the astration of primordial lithium most likely depends strongly on age, and that the lithium profile in that figure results from several unrelated phenomena, which occurred at different places and times over a period of some billion years.

\subsection{A Comment on the Ryan et al. Blue-Straggler Connection}

Against this background, we shall not consider further the ``blue straggler'' related lithium-poor-stars discussed by \citet{ryan01a, ryan02}.  There are relatively few of these objects\footnote{An estimate of the relative frequency of this type of star is afforded by the work of \citet{thorburn94}, who reported three of them in a sample of some 80 near-main-sequence stars having --3.0 $<$ [Fe/H] $<$ --1.5 and {\teff} $>$ 5500~K.}, and as noted by \citet{bonifacio12}, the soundness of the blue-straggler hypothesis notwithstanding, their restriction to relatively higher observed values of [Fe/H] suggests they can play only a minor role in our understanding of the complicated lithium abundance patterns at the earliest times.  In light of the potential relationship between this class of stars and ``blue stragglers" \citet{ryan02} advise: ``Such objects must be avoided in studies of the primordial Li abundance and in investigations into the way normal single stars process their initial Li".  One might wonder if near-main-sequence, C-rich, Li-weak stars with [Fe/H] $<$ --3.0 might be rapid rotators.  To our knowledge there exists little information on this possibility.  One exception is the analysis of HE~1327--2326 ([Fe/H] = --5.66, A(Li) = $<$0.70) by \citet{aoki06}, who report ``no clear excess (line) broadening by rotation ... with respect to Li-normal stars", from their careful high-resolution analysis (R = 60000).

\subsection{Three Population~II Subpopulations} {\label{sec:3pops}

Given the {\teff} and [Fe/H] of the stars under discussion here, we have implicitly assumed their membership of the Population II that resides in the Galactic halo (see \citealt{sandage86} and references therein for details of the concept of stellar populations).  In what follows, for heuristic purposes, we shall define three Population II subpopulations, which we name Pop~IIa, Pop~IIb, and Pop~IIab. Pop~IIa comprises C-rich stars with [Fe/H] $<$ --4.5 and [C/Fe] $\ge$ +0.7; Pop~IIb contains C-normal stars with [Fe/H] $>$ --4.2 and [C/Fe] $<$ +0.7;\footnote{We realize that some stars will be excluded by these definitions.  Fine tuning will be required as more/better data become available.} and in Section~\ref{sec:toymod} we shall introduce a toy model which postulates the formation of Pop~IIab stars as the result of the coalescence of material from Pop~IIa with that of Pop~IIb. 

\subsection{[Fe/H] $<$ --4.5 and [C/Fe] $>$ +0.7 (Pop~IIa)}
 
Only 15 stars are currently known that have [Fe/H] $<$ --4.5. We refer the reader to \citet{frebel15b}, and references therein, for an extensive literature concerning the chemical enrichment of the metal-free material that emerged following the Big Bang. A short list of topics includes minihalos, supermassive rapidly-rotating stars, mixing-and-fallback Type~II supernovae, Type~II supernovae with relativistic jets, and zero-metallicity, rotating, massive intermediate-mass stars. For the present study, we focus on the eight of those 15 stars with [Fe/H] $<$ --4.5 which lie near the main sequence. Six of them are characterized by large C abundances, with [C/Fe] {\simgt}~3.5. Their lithium abundances are in the range of $<$0.7 $<$ {\ali} $<$ 2.1.  Depending on their C abundances, low-metallicity stars have been grouped into the ``C-rich'' and ``C-normal'' objects. Regarding potential formation scenarios of these stars, \citet{chiaki17}, and others beforehand, e.g. \citep{frebel07}, discuss the role of carbon and its critical abundance of [C/Fe]$_{\rm b}$ = 2.30 that a gas needs to reach in order to cool enough through fine structure lines to lead to low-mass star formation. Below such a critical value, silicate grains would be the dominant coolant instead \citep{chiaki17, ji14}. 

The large lithium spread of $<$0.7 $<$ {\ali} $<$ 2.1, seen in panel (a) of Figure~\ref{fig:licfe_obs}, is at first sight a little surprising.  Why is it so large? What is its significance? \citet{meynet10} put it very succinctly in their study of CEMP stars: ``(Lithium) is completely destroyed in massive stars and also in AGB stars. Thus any mixing of such stellar ejecta with pristine interstellar material will increase the abundance of Li with respect to the abundance in the source material.''

\subsubsection{J0023+0307 and the Aguado et al. Suggestion}  

\citet{frebel19} and \citet{aguado19}, in their studies of J0023+0307\footnote{J0023+0307 = SDSS~J002314.00+030758.0} ([Fe/H] $<$ --5.8, {\ali} = 1.9, and [C/Fe] $>$ 3.8), discuss different sources of the ``pristine interstellar material''.  \citet{frebel19} conclude ``The dilution masses inferred here strongly suggest that J0023+0307 is a second-generation star formed by recollapse in a Population~III minihalo''.  \citet{aguado19} reach the conclusion that J0023+0307 and other stars in the range --6.0 $<$ [Fe/H] $<$ --2.5 provide an upper {\ali} envelope having ``a nearly constant value'' and that ``it is unlikely that such uniformity is the result of depletion processes in stars from a significantly higher initial Li abundance, but suggests instead a lower primordial production''.

The \citet{aguado19} suggestion brings to mind a potentially similar puzzle involving the red giant SM~0313--6708 $-$ the most [Fe/H]-poor star currently known ({\teff} = 5125~K, {\logg} = 2.3, [Fe/H] $<$ --7.3, [C/Fe] $>$ 4.9, and {\ali} = 0.7 \citep{keller14}).  \citet[see their Section 3.6 and Figure~6]{frebel15b} project the observed {\ali} backwards in time from the red giant branch to the value the star would have had on the main sequence, and report {\ali}$_{\rm MS}$ $\sim$~2.0. This is not unlike the {\ali} value of the upper Li envelope highlighted by \citet{aguado19} of {\ali} $\sim$ 2.0 (and 1.9 \citep{aguado19} and 1.7 \citep{frebel19} for J0023+0307 specifically). Clearly, more accurate observational data on the upper envelope of the lithium distribution, in the range of [Fe/H] $<$ --2.5, are needed to further investigate this important matter and to possibly settle the discussion of whether the primordial lithium abundance is perhaps environment dependent.

\subsection{\citet{fu15} ``From pre-main sequence to the Spite plateau'' (--4.2 $<$ [Fe/H] $<$ --1.5, Pop~IIb)}

A basic feature of Population~II is the Metallicity Distribution Function (MDF), which according to \citet[see also references therein]{dacosta19} ``has a power-law slope of $\Delta$(Log~N)/$\Delta$([Fe/H]) = 1.5 $\pm$ 0.1 dex per dex for --4.0 $\leq$ [Fe/H] $\leq$ --2.75, but appears to drop abruptly at [Fe/H] $\sim$ --4.2, in line with previous studies''. The number of C-rich stars with [Fe/H] $<$ --4.5 falls well above the extrapolation of this MDF.

The majority of Population~II is C-normal, with [C/Fe] $<$ 0.7.  For stars with [Fe/H] $<$ --1.0, however, C-rich objects ([C/Fe] $>$ 0.7) embrace the CEMP classification of \citet{beers05} and \citet{aoki07}, and the fraction of these stars increases as [Fe/H] decreases.  Below [Fe/H] = --3.0, C-rich stars comprise some 20 $-$ 40\% of Population II material \citep{yong13b, placco14}.

\citet{fu15} propose that the evolution of pre-main-sequence stars provides the site of the astration of primordial lithium.  They adopt modifications to standard theory that involve variations in the effects of envelope overshooting, residual mass accretion, EUV-photoevaporation, and main sequence diffusion and Li burning. Adopting conventional nuclear burning and microscopic diffusion along the main sequence, and beginning with the primordial lithium abundance, {\ali}$_{\rm P}$ = 2.72, they reproduce the Spite Plateau for stars in the metallicity range [M/H] = --3.2 $-$ --1.5. \citet{fu15} state: ``For our standard choice of parameters, stars with initial mass from m$_{0}$ = 0.62 to 0.80~M/M$_{\odot}$, nicely populate the Spite plateau (A(Li) $\sim$ 2.26).''  They also foreshadow ``the possibility to interpret the decrease of Li abundance in extremely metal-poor stars.''.  We shall assume that the \citet{fu15} hypothesis is the most likely stellar-evolutionary explanation of the Spite Plateau in this [Fe/H] abundance range, and for what follows remind the reader that we refer to C-normal stars with [Fe/H] $>$ --4.2 and [C/Fe] $<$ 0.7 as Pop~IIb.

One will also see the potential overabundance of riches within the current and previous subsections, given the conflict between the suggestions of \citet{fu15}, on the one hand, and \citet{aguado19}, on the other.  Said differently, there appears to be potential tension between stellar physics, on the one hand, and cosmological physics on the other.  We shall return to this possibility in Section~\ref{sec:summary}.

\subsection{A Toy Model to Explain CEMP-no Stars and the Li ``Meltdown'' in the Range --4.2 $<$ [Fe/H] $<$ --3.0 (Pop~IIab)} {\label{sec:toymod}

In \citet{norris19} we presented a toy model that sought to describe the formation of CEMP-no stars in the abundance range --4.0 {\simlt} [Fe/H] {\simlt}~--2.0 as the result of the coalescence of gas clouds from the two populations that followed the chemical enrichment by the first zero-heavy-element stars, namely the C-rich, ultra- and hyper-metal-poor (UMP and HMP) population with [Fe/H] $<$--4.5, on the one hand, and the C-normal, EMP halo stars having [Fe/H] {\simgt}~--4.0, on the other.  

\begin{figure*}
    \includegraphics[width=.90\hsize,angle=-90]{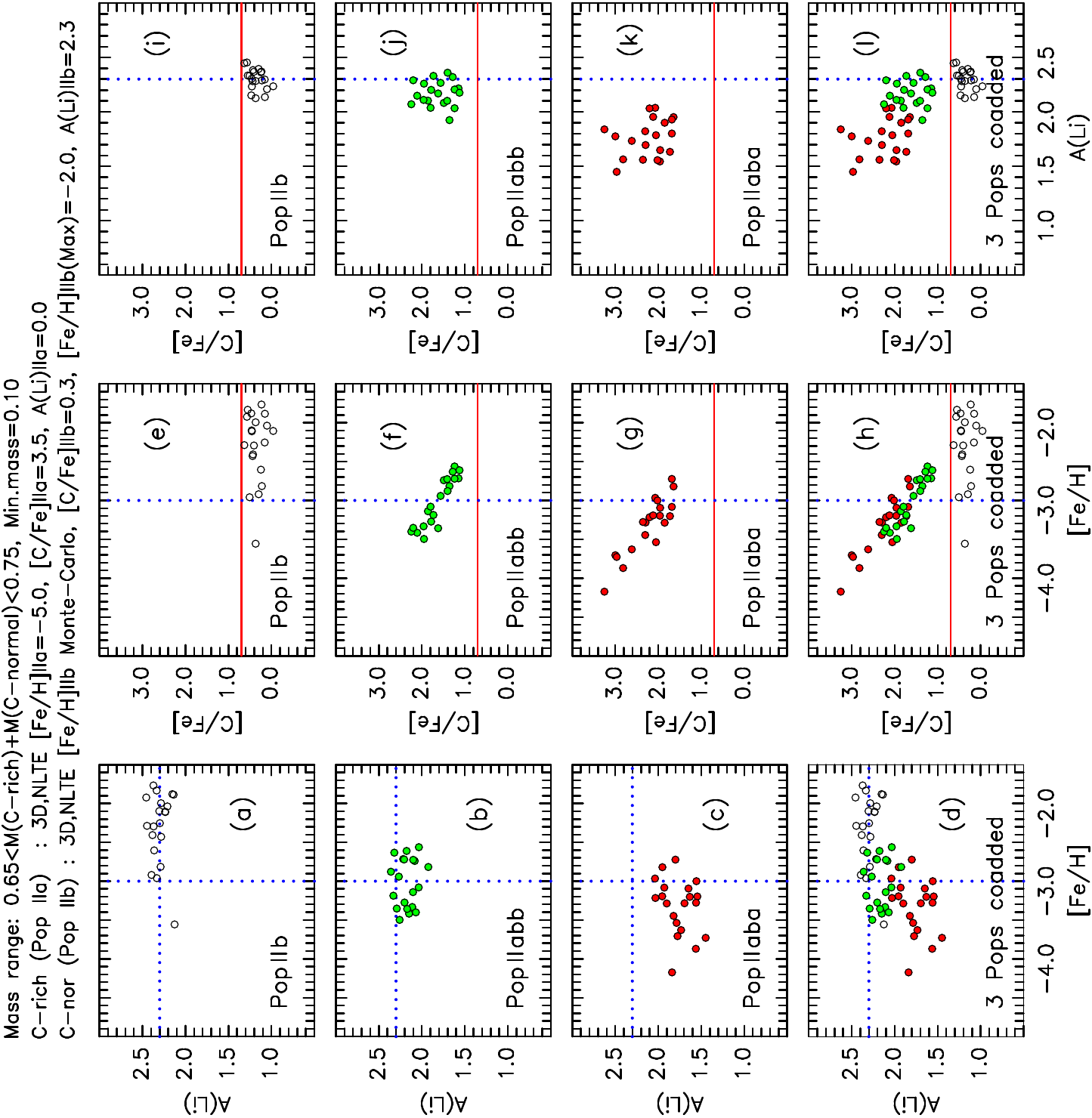}
    \caption{\small {(Left) {\ali}$_{\rm 1D,LTE}$ vs. [Fe/H]$_{\rm
          1D,LTE}$, (Middle) [C/Fe]$_{\rm 1D,LTE}$ vs. [Fe/H]$_{\rm
          1D,LTE}$, and (Right) [C/Fe]$_{\rm 1D,LTE}$ vs. {\ali}$_{\rm
          1D,LTE}$ for the toy-model three Population II
        subpopulations. From top to bottom the panels present Pop~IIb
        = C-normal halo, Pop~IIabb = C-normal dominated component,
        Pop~IIaba = C-rich dominated component, and the co-addition of
        Pop~IIb, Pop~IIabb, and Pop~IIaba. See text for discussion.}}
    \label{fig:toymod}
\end{figure*}

We refer the reader to \citet[Section 8]{norris19} and references therein for more details. Suffice it here to repeat the basic premise: ``The first generation of stars produced an initially carbon-rich environment in which further star formation proceeded along two principal pathways, one forming extremely carbon-rich objects (seen today as the C-rich stars with [Fe/H] {\simlt} --4.5; the minority population), the other (later) one comprising C-normal stars (seen today as the bulk of stars with [Fe/H] {\simgt} --4.0 $-$ the majority population).'' The simplicity of the model, and the uncertainty of the Fe and C abundance distributions and mass function of the HMP population notwithstanding, the model produced behaviour in the A(C) and [C/Fe] vs. [Fe/H] planes not unlike that seen in the \citet{yoon16} CEMP Groups I, II, and III.  In the present toy-model context, the C-rich, [Fe/H] $<$ --4.5 clouds are responsible for producing Pop~IIa, while the C-normal, [Fe/H] {\simgt} --4.0 clouds lead to the formation of Pop~IIb.  We now investigate to what extent this simple model might explain the apparent sub-Spite Plateau Li abundances of stars with [Fe/H] {\simgt} --4.2 seen in our Figures~\ref{fig:authors1}, \ref{fig:authors2}, and \ref{fig:lastly196alife}}.

The basic assumption of our toy model model is that the final mass of a putative composite star will be the sum of mass contributions arising from two independent star forming gas clouds. One cloud would have a C-normal chemical environment, the other would be a C-rich environment. To construct such a star, we then require gas mass contributions from each cloud with their respective chemical compositions. We draw each of said gas mass contributions at random from the Salpeter mass function but restricted to the range of 0.10 $\leq$ M/M$_{\odot}$ $\leq$ 0.75. To approximate the existence of 12-13\,Gyr metal-poor stars with [Fe/H] $<$ --2.0 now lying near the main-sequence turnoff, we then require that the composite star have a final mass in the range of 0.65 $\leq$ M/M$_{\odot}$ $\leq$ 0.75.

Having fixed the abundances of each of the randomly chosen contributing masses, we then determine the chemical abundances of Fe, C, and Li of the composite star. Details are further discussed below.  

We note that a basic shortcoming of our toy model pertains to the lower limit on the lesser of the two gas mass contributions to a composite star. This 0.1{\msun} limit is not well constrained, and smaller contributions may occur in reality. The distribution of relative contributions is also unknown. Consequently, computational conveniences -- the 0.1{\msun} lower limit and the Salpeter IMF -- have been adopted. Within these caveats some insights may still be gained and some progress may be possible.

In addition, the model is agnostic to  the details concerning the formation process and location of the putative composite star. However, the idea that two gas mass reservoirs contribute to the making of a star is not new. Recently, a version of it was explored by \citet{smith15} who modeled external enrichment by a neighboring minihalo. The basic scenario follows a minihalo in the early universe that becomes chemically enriched by a neighboring minihalo whose massive supernova injects metals at high speed into it. Such a scenario has also been suggested to explain the different populations of C-rich and C-normal stars \citep{ezzeddine19} assuming that significant amounts of carbon produced by a hypernova were acquired by an Fe-poor minihalo.  In these cases the external enrichment scenario is naturally limited to very early, low-metallicity environments, consistent with the approach adopted in the present study.
 
\subsubsection{\it The Composite Pop~IIab} {\label{sec:popiiab}

In Section~\ref{sec:3pops}, we defined three Population II subpopulations: (1) Pop~IIa (the C-rich stars with [Fe/H] $<$ --4.5), (2) Pop~IIb (the C-normal stars with [Fe/H] $>$ --4.2), and (3) Pop~IIab (composites of material of these two subpopulations).

Assuming that the toy-model predictions would best represent 3D,NLTE observational values, rather than 1D,LTE, we re-define our model parameters accordingly.  Given the small number of observed stars with [Fe/H} $<$ --4.5, we have adopted representative population parameters of [Fe/H]$_{\rm 3D,NLTE}$ = --5.00, {\ali}$_{\rm 3D,NLTE}$ = 0.00, and [C/Fe]$_{\rm 3D,NLTE}$ = 3.50; and for the C-normal Pop~IIb we adopt a modified \citet{yong13b} MDF over the range --4.0 $<$ [Fe/H]$_{\rm 3D,NLTE}$ $<$ --2.0, together with {\ali}$_{\rm 3D,NLTE}$ = 2.30, and [C/Fe]$_{\rm 3D,NLTE}$ = 0.30.  For each M$_{\rm C-rich}$, M$_{\rm C-normal}$ combination of the composite star, the Pop~IIab [Fe/H]$_{\rm 3D,NLTE}$ and [C/Fe]$_{\rm 3D,NLTE}$ values then follow. We use transformations based on \citet{norris19} to obtain 1D,LTE predictions{\footnote{[Fe/H]$_{\rm 1D,LTE}$ = [Fe/H]$_{\rm 3D,NLTE}$ -- 0.35 and [C/H]$_{\rm 1D,LTE}$ = [C/H]$_{\rm 3D,NLTE}$ -- 0.17[Fe/H]$_{\rm 3D,NLTE}$ -- 0.09.}} for comparison with the observational results in Figure~\ref{fig:licfe_obs}.

It is these composite stars that we refer to above as Pop~IIab.  By definition, Pop~IIab has two components: in one, which we shall call Pop~IIaba, the larger fraction of each star comprises (C-rich) Pop~IIa material; and in the second, Pop~IIabb, the larger fraction consists of (C-normal) Pop~IIb material.  The significance of the subdivision is that Pop~IIaba will on average have higher [C/Fe] and lower {\ali} values than those of Pop~IIabb, and vice-versa.  This difference drives the two subpopulations apart in the {\ali} and [C/Fe] vs. [Fe/H] planes.

Figure~\ref{fig:toymod} presents this example of the behaviour of the three subpopulations Pop~IIb, Pop~IIaba, and Pop~IIabb in the {\ali}$_{\rm 1D,LTE}$ and [C/Fe]$_{\rm 1D,LTE}$ vs [Fe/H]$_{\rm 1D,LTE}$ planes, and that of [C/Fe]$_{\rm 1D,LTE}$ vs {\ali}$_{\rm 1D,LTE}$.  (We have added Gaussian errors of 0.05, 0.10, and 0.15 dex to the model {\ali}, [Fe/H], and [C/Fe] values, respectively). While we have no means of determining the relative numbers of the three subpopulations, by way of example we have plotted the positions of 20 toy-model stars in each of the upper three rows of the figure, together with their co-addition in the bottom row.  In the figure, the left column contains {\ali}$_{\rm 1D,LTE}$ vs. [Fe/H]$_{\rm 1D,LTE}$, the middle column presents {\ali}$_{\rm 1D,LTE}$ vs. [C/Fe]$_{\rm 1D,LTE}$, and on the right we plot [C/Fe]$_{\rm 1D,LTE}$ vs. {\ali}$_{\rm 1D,LTE}$.  From top to bottom, each column contains Pop~IIb (C-normal), Pop~IIabb (composite stars dominated by (C-normal) Pop~IIb), Pop~IIaba (composites dominated by (C-rich) Pop~IIa), and Pop~IIb + Pop~IIabb + Pop~IIaba (their co-addition), respectively.

We are encouraged to suggest that there is a significant similarity between the {\ali}$_{\rm 1D,LTE}$ vs. [Fe/H]$_{\rm 1D,LTE}$ co-added panel (d) in Figure~\ref{fig:toymod} and the observed abundance distributions in Figure~\ref{fig:lastly196alife}.  It is perhaps all that one might hope for, insofar as the observed sample is also very statistically incomplete, as the result of selection effects.  Perhaps the most interesting aspect of the toy model, however, is the prediction that Pop~IIaba and Pop~IIabb will occupy somewhat different regions of the diagram in the vicinity of [Fe/H] = --3.0.  In particular, concerning the report by \citet{matsuno17b} that there are CEMP-no stars with {\ali} = 2.1 $-$ 2.2 (see our Table~\ref{tab:tab6}), perhaps these stars belong to the toymodel composite Pop~IIabb, in which there is a dominance of Pop~IIb material.

There also appears to be a significant similarity between the [C/Fe]$_{\rm 1D,LTE}$ vs. {\ali}$_{\rm 1D,LTE}$ panels (i) $-$ (l) in Figure~\ref{fig:toymod}, on the one hand, and the observed abundance distributions in Figure~\ref{fig:licfe_obs}, on the other.

\subsubsection{Caveat Emptor}

These selected favorable comparisons notwithstanding, we hasten to acknowledge that the toy model is at best only suggestive.  There are many free parameters in our presentation $-$ for example, the single C-rich Pop~IIa component represented by only one set of parameters\footnote{See \citet[Figures 9 $-$ 11]{norris19} for changes that occur when different Pop~IIa parameters are adopted.}, the mass ranges of the two clouds that coalesce (also the assumption that there is no mass loss when they combine), and the [Fe/H] range of the Pop~IIb clouds.  Our aim here is to show that one can find a set of parameters that appears to reproduce the observational data.  We hope that other authors, theoretically inclined, will accept the challenge to investigate in more detail the concept we have proposed.

\subsection {The Sloping of the Spite Plateau} \label{sec:slope}

We conclude our discussion by recalling that while in Section~\ref{sec:intro} we suggested there are five lithium problems, we have so far not discussed one of them $-$ the slope of the Li plateau in the ({\ali}, [Fe/H]) -- plane in the range --3.5 {\simlt} [Fe/H] {\simlt} --1.5 (other than to quantify its value in Figures~\ref{fig:authors1} and \ref{fig:authors2}).  Examination of panels (a) $-$ (d) of Figure~\ref{fig:toymod} suggests that a thin and horizontal plateau of Pop~IIb stars merging with Pop~IIab stars, at [Fe/H] $\sim$~--3.0 $-$ --2.5, could provide a simple explanation.

In Section~\ref{sec:3steps}, we presented the average values and dispersions of {\ali} for stars with {\teff} $>$ 6000~K, and noted that for those in the range --2.2 $<$ [Fe/H] $<$ --1.7 (within the \citet{spite82} discovery range) the plateau appears horizonal in the {\ali} vs. {\teff} regime of the ``Literature'' and ``Corrected-Literature'' panels.  To investigate this suggestion further we considered the lithium parameters of the Spite Plateau in the range --2.0 $<$ [Fe/H] $<$ --1.0, using the data for the 36 stars in the intersection of the \citet{asplund06} and \citet{melendez10} samples, which we regard to be of the highest quality within this abundance range.  We find $\langle${\ali}$\rangle$ = 2.322 $\pm$ 0.013, with $\sigma$({\ali}) = 0.078, resulting from 35 stars (one 3$\sigma$ outlier having been omitted).  We would also draw the reader's attention to the fact that CEMP-no stars are found only below [Fe/H] = --2.0 (\citealt{aoki10}, \citealt{norris13}), supporting the above suggestion of the merger of two subpopulations.  

\section{Summary} \label{sec:summary}

We have collated and discussed literature lithium abundances of some 200 near-main-sequence stars with {\teff} $>$ 5800~K and [Fe/H] $<$ --1.5.  Three different approaches to the data were explored: (1) adopt ``Literature'' values $-$ the data remain unchanged; (2) homogenise the data to obtain ``Corrected-Literature'' values $-$ moving the data onto a single stellar temperature (IRFM) scale, and (3) determine ``{\it ab initio} IRFM {\tirfm}'' values $-$ Infrared Flux Method temperatures from literature infrared colours.  

We then examined {\ali} as a function of {\teff}, [Fe/H], and [C/Fe].  In Section~\ref{sec:intro}, five aspects of the complicated distribution of stars in the ({\ali}, [Fe/H]) plane were identified which challenge insight into the status of lithium abundances at the earliest times. We conclude here with a summary of the potential implications of these five problems.

\begin{itemize}

\item
Very Low {\ali} Values in the Group of Stars Related to ``Blue Stragglers''\\
\vspace{-0.3cm}

This is a minority $\sim$5\% population, noted by \citet{bonifacio12} to be restricted to relatively high [Fe/H] values, suggesting these stars are not relevant to lithium abundance patterns at the earliest times.\\

\item
Lithium in the Most Iron-poor Stars ([Fe/H] $<$ --4.5)\\
\vspace{-0.3cm}

These are the first stars following Population III and the oldest
Population II objects (which we designate Pop~IIa) $-$ lithium-poor
($<$0.5 $<$ {\ali} $<$ 2.1) and extremely carbon-rich (most having
[C/Fe] {\simgt}~3.5). They formed earliest at the epoch when carbon grains
were the dominant gas coolant.  \citet{aguado19} suggest there are
stars in the range --6.0 $<$ [Fe/H] $<$ --2.5 that provide an upper
{\ali} envelope at ``a nearly constant value'' of {\ali} $\sim$ 2.0.
They argue this value is the Primordial Lithium Abundance.\\

\item
Lithium ``Meltdown'' in Stars in the Range --4.2 $<$ [Fe/H] $<$ --3.0\\
\vspace{-0.3cm}

The second and later pathway of star formation, in which silicate grains are the dominant gas coolants, leads to C-normal population stars, seen today as the bulk of stars with [Fe/H] {\simgt} --4.2 (which we call Pop IIb). We presented a toy model in which gas from Pop~IIa and Pop~IIb combine to form Population II stars that we call Pop~IIab, which form a large part of the ``meltdown'', and in which there is an anticorrelation between lithium and carbon.  That is, the lithium ``meltdown'' is accompanied by carbon enrichment, with [C/Fe] values as large [C/Fe] = +2.0 to +3.0.\\

\item   
The Slope of the Spite Plateau in the {\ali} vs. [Fe/H] Plane\\
\vspace{-0.3cm}

Over the range --3.5 $<$ [Fe/H] $<$ --1.5 the sloping of the Spite Plateau appears to be caused by the merging of two subpopulations.  The first constitutes a plateau of C-normal Pop~IIb stars which has $\langle${\ali}$\rangle$ = 2.322 $\pm$ 0.013 over the range --2.0 $<$ [Fe/H] $<$ --1.0.  The second is the subpopulation of C-rich Pop~IIab stars.  The latter resulted from the combination of C-rich and strongly Li-depleted (Pop IIa) and C-normal (Pop IIb) material.\\

\item
The Primordial Lithium Problem\\
\vspace{-0.3cm}

\citet{fu15} propose pre-main-sequence evolution as the site of the astration of primordial lithium.  Their model explains the Spite Plateau in the metallicity range [M/H] = --3.2 $-$ --1.5, and for their ``standard choice of parameters, stars with initial mass from m$_{0}$ = 0.62 to 0.80~M/M$_{\odot}$, nicely populate the Spite plateau (A(Li) $\sim$ 2.26).''

According to \citet{cyburt08}, the primordial lithium abundance is {\ali}$_{\rm P}$ = 2.72 $\pm$ 0.06, compared with an observed value {\ali} = 2.09 $\pm$ 0.03.  Our updated $\langle${\ali}$\rangle$ value is 2.322 $\pm$ 0.013).  (Note that the more recent \citet{planck20} CMB results ``do not discuss other light elements, such as ... lithium ... and do not shed any further light on earlier CMB experiments''.)

In Section~\ref{sec:discussion} we cited (1) the suggestion of \citet{aguado19} that stars in the range --6.0 $<$ [Fe/H] $<$ --2.5 provide an upper lithium abundance envelope at nearly constant value {\ali} $\sim$ 2.0, ``suggesting  a lower primordial production'' than the currently accepted value; and (2) the most [Fe/H]-poor star currently known $-$ the red giant SM~0313--6708, with [Fe/H] $<$ --7.3, {\logg} = 2.3, [C/Fe] $>$ 4.9, and {\ali} = 0.7 \citep{keller14} $-$ is suggested by \citet{frebel15b} to have had {\ali} $\sim$~2.0 when on the main sequence.

Given this tension between lithium abundances based on stellar and cosmological endeavors, it is then of considerable interest that \citet{riess19} report conflicting Hubble Constants (H$_{0}$) of 74.03 $\pm$ 1.42 km\,s$^{-1}$\,Mpc$^{-1}$ (from local observations) compared with 67.4 $\pm$ 0.5 km\,s$^{-1}$\,Mpc$^{-1}$ (inferred from Planck CMB \citep{planck20} plus Big Bang $\Lambda$CDM) $-$ a difference of 6.6 $\pm$ 1.5 km\,s$^{-1}$\,Mpc$^{-1}$.  A similar, independent, conclusion has also been reached by \citet{freedman19}.  That is, there are tensions between stellar- and cosmological-based estimates of both the primordial lithium abundance and the Hubble Constant.  

With this in mind we noted in Section~\ref{sec:discussion} that more observational data on the upper envelope of the lithium distribution in the abundance range --6.0 $<$ [Fe/H] $<$ --2.5 are needed to address the lithium aspect of these issues.

\end{itemize}

\begin{figure*}

    \includegraphics[width=.70\hsize]{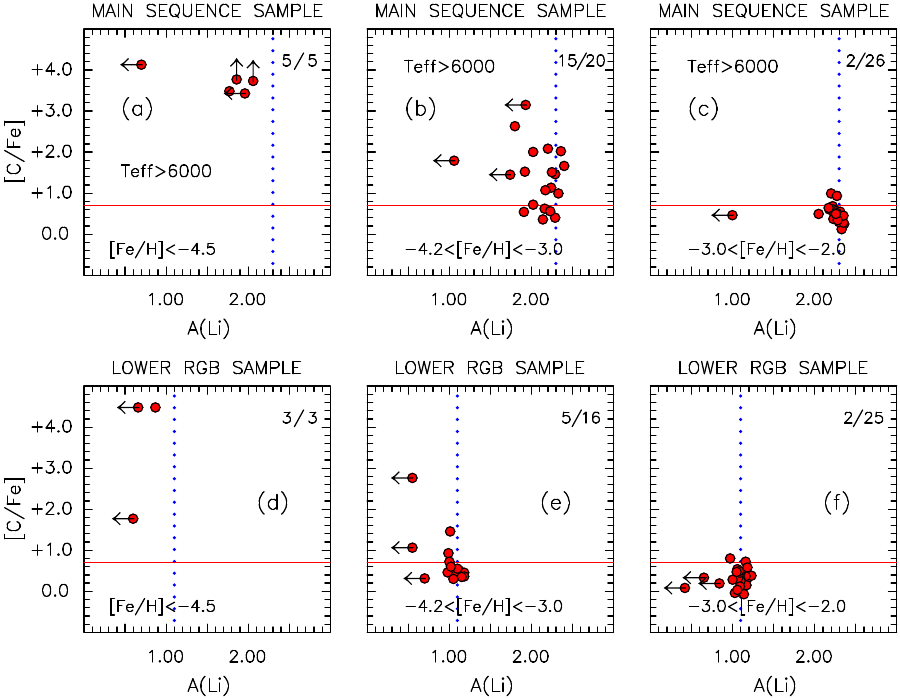}
    \caption{\small  [C/Fe] vs. {\ali} for stars with (left) [Fe/H] $<$
          --4.5, (middle) --4.2 $<$ [Fe/H] $\leq$ --3.0, and (right) --3.0 $<$ [Fe/H] $\leq$ --2.0.  The upper and lower panels present data for main sequence ({\teff} $>$ 6000~K) and LRGB stars, respectively.  At top right of each panel the two numbers form the ratio of the number of C-rich stars to the total number of stars (excluding the stars with upper limits). The black vertical lines represent Primordial Lithium ({\ali}$_{\rm P}$ = 2.72), the red line at {\ali} = 0.7 separates C-weak and C-strong stars, and the blue dotted lines are for reference purposes.  See text for details.}
    \label{fig:appendix}
\end{figure*}

{\bf{Appendix}} \\

{\bf{A Comparison between Main-Sequence and Lower Red Giant Branch Lithium Abundance Distributions}} \\

The final entry in our Table 1 (Milestones in the Study of Lithium Abundances) is ``Discovery of a thin lithium plateau among metal-poor red giant branch stars'' of \citet{mucciarelli22}.  In this work, its authors report: ``The Lower RGB (LRGB) stars display an A(Li) distribution that is clearly different from that of the dwarfs, without signatures of a meltdown and with two distinct components: (a) a thin A(Li) plateau with an average A(Li) = 1.09 $\pm$ 0.01 dex ($\sigma$= 0.07 dex) and (b) a small fraction of Li-poor stars with A(Li) lower than $\sim$~0.7 dex''. 

To investigate the apparent inconsistency between the relative simplicity of the LRGB Li distribution compared to the five lithium problems shown schematically in our Figure 1, we compare the values presented in the \citet{mucciarelli22} Table 1 and Figure 2 (their A(Li) vs. [Fe/H] and {\teff}) with the data in our Table~7 and Figure~10.  A basic difference between the two approaches is that while Figure 2 of \citet{mucciarelli22} centres on A(Li) vs. [Fe/H], our Figure 10 examines the inter-relationship between [Fe/H], [C/Fe], and A(Li).  Carbon and iron remain essentially unchanged in the evolutionary transition from main sequence to LRGB{\footnote{\citet{placco14} show that [C/H] corrections are always less than 0.06 dex for RGB stars having {log g} $>$~2.0, the lower limit of the \citet{mucciarelli22} sample.}.  We address the possibility that carbon may be relevant for an understanding of the apparent difference between the above two systematisations.

Our first step is to seek to minimize the differences in selection effects between the two approaches.  In particular, our analysis included only stars with [Fe/H] $\leq$~--2.0 as the upper limit of our sample in Figure 10.  We therefore adopt that limit here.  Then, given that CEMP-s stars were excluded from the main-sequence group (see our Section 2) the three CEMP stars of \citet{mucciarelli22} which are also CEMP-s $-$ CS~29495-042, HE~0207--1423, and HE~1005--1439 $-$ (see \citealt{masseron10}, \citealt{yong13a}, and \citealt{goswami22}, respectively) are excluded from the present comparison. Finally, following \citet{aoki07}, we define C-rich stars as those having [C/Fe] $>$ 0.7. With the above restrictions, the \citet{mucciarelli22} sample size reduces from 58 to 44 stars, while the number of main-sequence stars in our Table 7 that have {\teff} $>$ 6000~K changes from 78 to 70.  A significant difference between the two datasets is that while all stars in the LRGB group have observed [C/Fe] values, this is not the case for the main-sequence stars, for which 25 have only upper limits.  Against this background, we shall initially discuss only those stars with {\teff} $>$ 6000~K in which carbon has been detected. 

Figure~\ref{fig:appendix} presents the results for the two datasets, with the main-sequence sample in the upper three panels and the LRGB stars in the lower three.  The format is similar to that in our Figure 10: [C/Fe] vs. A(Li) is plotted for three [Fe/H] regimes, increasing from left to right: (left) [Fe/H] $<$ --4.5, (middle)--4.2 $<$ [Fe/H] $\leq$ --3.0, and (right) --3.0 $<$ [Fe/H] $\leq$ --2.0 for both data sets\footnote{Given the [Fe/H] boundaries adopted in our Figure 10, we have made a small change to the \citet{mucciarelli22} data for CS~0557--4840, for which they report [Fe/H] = --4.44. For convenience we have adopted [Fe/H] = 4.51, which places it in the most metal-poor category; this may be acceptable to some extent given that we have previously derived [Fe/H] = --4.75 for this star (\citealt{norris07, norris18}).}.  In all panels the red line represents [C/Fe] = 0.70, above which stars are C-rich; in the upper panels, as in Figure 10, the dotted blue line represents A(Li) = 2.3, while in the three lower panels A(Li) = 1.09, adopted following \citet{mucciarelli22}.  

The numbers in the top right of each panel in the figure represent the ratio of the number of C-rich stars to the total number of stars (excluding those with only upper limiting [C/Fe] values). In the leftmost panels all stars are C-rich, while in the rightmost the ratios contain mainly C-normal stars with ratios close to 0.08. The two middle panels are more interesting, both exhibiting significantly large spreads in [C/Fe] values.  The ratios of C-rich stars to all stars are very different - for the main-sequence stars the ratio is 0.75 while for the LRGB stars it is 0.31 - but these figures disguise very different selection biases dictated by the temperature- and pressure-differences in the carbon detection limits for (warm) main sequence stars and (cooler) LRGB stars. The LRGB ratio of 0.31 for the C-rich fraction is in excellent agreement with the ratios (fractions) of $\sim$0.2 -- 0.4 for CEMP-no stars in the abundance range --3.8 $<$ [Fe/H] $<$ --3.0 reported by \citet[their Figure 7]{yong13b} (see also \citealt{placco14}, and references therein).

For the main sequence group, we have excluded a significant number of stars that have only upper limit estimates for [C/Fe]. In our Figure 10 one sees that the total number of stars in panel (b) (--4.2 $<$ [Fe/H] $\leq$ --3.0) (including  those with only upper [C/Fe] limits) is 29. If one were to assume that those with upper limit are C-normal, the main-sequence ratio would become 15/29 = 0.52 $-$ not too far from the mark.  The likely explanation of the large ratio for the main-sequence sample is that the data selection procedure adopted here is biased towards stars having CEMP-no characteristics and/or large data samples (see for example \citet{barklem05}, \citet{jacobson15}, \citet{li15}, \citet{placco16}, \citet{siva06}, and \citet{spite13}. 

Against this background, and in particular the presence of significant numbers of stars having both low lithium and high carbon in the abundance range --4.2 $<$ [Fe/H] $\leq$ --3.0 in both main-sequence and LRGB samples, we are of the view that the same physical process most likely occurred in the two data sets. \\ 

}

\noindent  {\bf{Acknowledgements}} \\

J.E.N. acknowledges Constantine Deliyannis for his insistence in the late 1990s that ``there are cases of well-observed stars ... which cannot have the same Li abundance'' (see \citealt{ryan96}), as proved to be the case.  We thank Piercarlo Bonificacio for his comments on the manuscript, in particular for alerting us to the work of, and his comments on, \citet{mucciarelli22}.  Parts of this research were supported by the Australian Research Council Centre of Excellence for All Sky Astrophysics in 3 Dimensions (ASTRO 3D), through project number CE170100013.  A.F. acknowledges support from NSF grant AST-1716251.  This work has made use of data from the European Space Agency (ESA) mission {\it Gaia} (\url{https://www.cosmos.esa.int/gaia}), processed by the {\it Gaia} Data Processing and Analysis Consortium (DPAC, \url{https://www.cosmos.esa.int/web/gaia/dpac/consortium}). Funding for the DPAC has been provided by national institutions, in particular the institutions participating in the {\it Gaia} Multilateral Agreement.

\section{Data Availability} 

The data underlying this article will be shared on reasonable request to the corresponding author.



\bibliographystyle{mnras}
\bibliography{bibdesk_dy} 
\bsp	
\label{lastpage}
\end{document}